%% file: main.tex
\documentclass[manuscript]{acmart}

\AtBeginDocument{%
  }

\setcopyright{acmlicensed}
\acmJournal{CSUR}
\acmYear{2024} \acmVolume{1} \acmNumber{1} \acmArticle{1} \acmMonth{1}\acmDOI{10.1145/3653974}

\usepackage{multirow}
\usepackage{float}
\usepackage{enumitem}
\usepackage{tabularx}
\usepackage{url}
\usepackage{subfigure}

\usepackage{amsfonts,amsmath,amsthm}
\usepackage{ulem}

\usepackage{colortbl}
\usepackage{float}
\usepackage{amsfonts}
\usepackage{pifont}

\usepackage{booktabs} 
\usepackage{fdsymbol}
\usepackage{adjustbox}
\usepackage{wasysym}
\newcolumntype{R}[2]{%
    >{\adjustbox{angle=#1,lap=\width-(#2)}\bgroup}%
    l%
    <{\egroup}%
}
\newcommand*\rota{\multicolumn{1}{R{90}{1em}}}

\begin{document}

\title{Recovery from Adversarial Attacks in Cyber-physical Systems: Shallow, Deep and Exploratory Works}



\author{Pengyuan Lu}
\affiliation{%
  \institution{University of Pennsylvania}
  \city{Philadelphia}
  \state{Pennsylvania}
  \country{USA}
}
\email{pelu@seas.upenn.edu}

\author{Lin Zhang}
\affiliation{%
  \institution{University of Pennsylvania}
  \city{Philadelphia}
  \state{Pennsylvania}
  \country{USA}
}
\email{cpsec@seas.upenn.edu}

\author{Mengyu Liu}
\affiliation{%
  \institution{University of Notre Dame}
  \city{Notre Dame}
  \state{Indiana}
  \country{USA}
}
\email{mliu71@syr.edu}

\author{Kaustubh Sridhar}
\affiliation{%
  \institution{University of Pennsylvania}
  \city{Philadelphia}
  \state{Pennsylvania}
  \country{USA}
}
\email{ksridhar@seas.upenn.edu}

\author{Oleg Sokolsky}
\affiliation{%
  \institution{University of Pennsylvania}
  \city{Philadelphia}
  \state{Pennsylvania}
  \country{USA}
}
\email{sokolsky@seas.upenn.edu}

\author{Fanxin Kong}
\affiliation{%
  \institution{University of Notre Dame}
  \city{Notre Dame}
  \state{Indiana}
  \country{USA}
}
\email{fkong@nd.edu}

\author{Insup Lee}
\affiliation{%
  \institution{University of Pennsylvania}
  \city{Philadelphia}
  \state{Pennsylvania}
  \country{USA}
}
\email{lee@seas.upenn.edu}

\renewcommand{\shortauthors}{Lu et al.}

\newcommand\ML[1]{{\color{blue}$\clubsuit${\color{blue} ML: #1}}} 
\newcommand\LZ[1]{{\color{brown}$\spadesuit${\color{brown} LZ: #1}}} 
\newcommand\KS[1]{{\color{blue}$\heartsuit${\color{blue} KS: #1}}} 
\newcommand\EL[1]{{\color{blue}$\diamondsuit${\color{blue} EL: #1}}} 
\newcommand\FK[1]{{\color{magenta}$\diamondsuit${\color{magenta} FK: #1}}} 
\newcommand\OS[1]{{\color{orange}$\clubsuit${\color{orange} OS: #1}}} 
\newcommand\IL[1]{{\color{red}$\spadesuit${\color{red} IL: #1}}} 

\newcommand\edit[1]{{\color{black}#1}}
\newcommand\editcr[1]{{\color{blue}#1}}


\begin{abstract}
\input{sections/0_abstract}
\end{abstract}



\maketitle

\input{sections/1_intro}
\input{sections/2_preliminary}
\input{sections/3_shallow}
\input{sections/4_deep}\\
\input{sections/5_exploratory}
\input{sections/6_discussion}
\input{sections/6_1_challenges_and_future}
\input{sections/7_conclusion}
\input{sections/acknowledgement}

\input{references}

\end{document}

%% file: sections/0_abstract.tex
Cyber-physical systems (CPS) have experienced rapid growth in recent decades. However, like any other computer-based systems, malicious attacks evolve mutually, driving CPS to undesirable physical states and potentially causing catastrophes. Although the current state-of-the-art is well aware of this issue, the majority of researchers have not focused on CPS recovery, the procedure we defined as restoring a CPS's physical state back to a target condition under adversarial attacks. To call for attention on CPS recovery and identify existing efforts, we have surveyed a total of 30 relevant papers. We identify a major partition of the proposed recovery strategies: shallow recovery vs. deep recovery, where the former does not use a dedicated recovery controller while the latter does. Additionally, we surveyed exploratory research on topics that facilitate recovery. From these publications, we discuss the current state-of-the-art of CPS recovery, with respect to applications, attack type, attack surfaces and system dynamics. Then, we identify untouched sub-domains in this field and suggest possible future directions for researchers.

%% file: sections/1_intro.tex
\section{Introduction}
\label{sec:introduction}

Cyber-physical systems (CPSs) integrate control, computing and sensing through physical components, and have been rapidly sprouting over the past years \cite{cpsKnowledgeArea2019}. This new type of paradigm requires technologies from various fields such as embedded systems, wireless networks, and control theory, etc. So far, the technology of CPS has boosted various applications to sprout, including unmanned autonomous vehicles (UAVs), smart grids, and robotic arms. Market data shows that the CPS is in a period of rapid growth. In 2013, the worldwide CPS market is anticipated to be worth \$44 billion dollars. The market for CPS is anticipated to increase by 7.6\% from 2013 to 2021. Recent years have seen a rise in the use of technology and the incorporation of Artificial Intelligence (AI) across several sectors in North America, both of which have contributed to the region's development. The global CPS market size is expected to grow from 86 billion dollars in 2022 to 137 billion dollars by 2028~\cite{fks}.

Unfortunately, malicious attacks mutually evolve with the development in CPS and potentially lead to catastrophes \cite{kong2018cyber, he2020exploring, kong2019state}. Similar to traditional software systems, CPSs are burdened by failures caused by adversaries. These attacks can come in various forms with catastrophic consequences. For example, the cyber worm Stuxnet penetrates programmable logical controllers in industrial plants and has been estimated to have infected over 60,000 computers around the globe upon its discovery in 2010 \cite{stuxnet2011}. 
Another example is transduction attacks by speakers, which can interfere the gyroscope readings in drones, leading to undesired flight trajectories and even crashes \cite{droneSpeakerAttack2015}.
The defense against a large variety of attacks have called attention of researchers, since CPSs include numerous life-critical applications such as autonomous vehicles and medical devices \cite{cpsKnowledgeArea2019}, and any failure due to attacks may entails devastating results such as property damages or lethal consequences. 

Consequently, an objective in CPS research is to counteract the adversarial attacks, by restoring desirable target \textbf{physical states} after an attack is landed. For example, a target state for a power grid can be outputting voltages within a safe range, and for an autonomous vehicle can be parking on the shoulder of the road. In this paper, we denote this procedure of restoration as \textbf{recovery}, a term borrowed from previous literature \cite{kong2018cyber,zhang2020lpRecovery,ma2020rmpc}. To formalize this procedure, we first define a CPS's target  states as follows.  
\begin{definition}[Target physical states]
    \label{def:target_states}
    Given a physical state space $\mathcal{S}$, we denote the physical state of a CPS at time $t$ as $s_t \in \mathcal{S}$. A physical state $s_t$ is a target physical state with respect to a predicate $P$ iff it satisfies $P$. 
\end{definition}
Generally, states can be divided into physical states and cyber states, with the latter denoting the states on cyber variables. 
In this paper, the recovery objective is to restore predefined physical states only, and we will use the term ``states" for physical states from now on. The reason for focusing on physical states is that they relate to damages in our physical world, leading to severe casualties such as property destruction, injuries or even lethal consequences.

Notice that this predicate $P$ in Definition \ref{def:target_states} specifies the target condition of states. For example, an autonomous vehicle may have a 3-dimensional physical state space $\mathcal{S} = [-100 m, 100 m] \times [-10 m, 10 m] \times [0 m/s, 20 m/s]$. Here, the first two dimensions are the vehicle's allowable positions in a coordinate, and the last dimension is its allowable speeds. Then, if the shoulder of the road is a position region $[-100 m, 100 m] \times [9 m, 10 m]$, the target condition ``parking on the shoulder of the road" can be specified as $P = s_t \in [-100 m, 100 m] \times [9 m, 10 m] \times \{0 m/s\}$. So far, researchers have expressed this target predicate $P$ in various formal languages, including linear temporal logic (LTL) \cite{pnueli1977temporal}, metric temporal logic (MTL) \cite{koymans1990specifying} and signal temporal logic (STL) \cite{maler2004monitoring}. From the previous definition, we now formalize the procedure of recovery as follows.

\begin{definition}[Recovery]
    \label{def:recovery}
    A CPS recovery, or simply recovery, is an online procedure that starts at some time $t_1$, where the system's state $s_{t_1}$ is not a target state with respect to a predefined predicate $P$. The procedure aims to produce a target state $s_{t_2}$ at a later time $t_2 > t_1$ with respect to $P$. Optionally, an additional objective is to maintain the satisfaction of $P$ from $t_2$ until some $t_3 > t_2$.
\end{definition}

So far, researchers have proposed various methods to design recovery algorithms that meet Definition \ref{def:recovery}. There exist multiple ways to categorize these solutions. For example, one partition is whether the procedure requires a dedicated recovery controller, which serves for recovery purpose only and splits the CPS into normal vs. recovery modes. Another partition is by different adversarial scenarios, such as different attack surfaces like sensors, actuators and communication channels. What is more, based on the type of application and scenario, people have different recovery targets, i.e. the condition $P$ to be satisfied. For example, $P$ can be a condition of safety, stability, liveness, or other desirable properties.


\edit{In this survey, we partition the papers by their recovery methods. This decision is made from a practical standpoint, so that readers can easily tell what tools have been adopted in the state-of-the-art, under what assumptions, and with what effects. We categorize the papers by two major categories: methods that do not leverage a controller for recovery purpose, and methods that do so. The reason for this categorization is because there exist various useful tools in the control domain. By reducing the recovery problem to a control problem, these tools can be applied to solve the recovery problem.} We refer to these two categories of recovery methods as \textbf{shallow and deep recovery}, respectively, defined as follows.

\begin{figure}[!htb]
    \centering
    \includegraphics[width=0.8\textwidth]{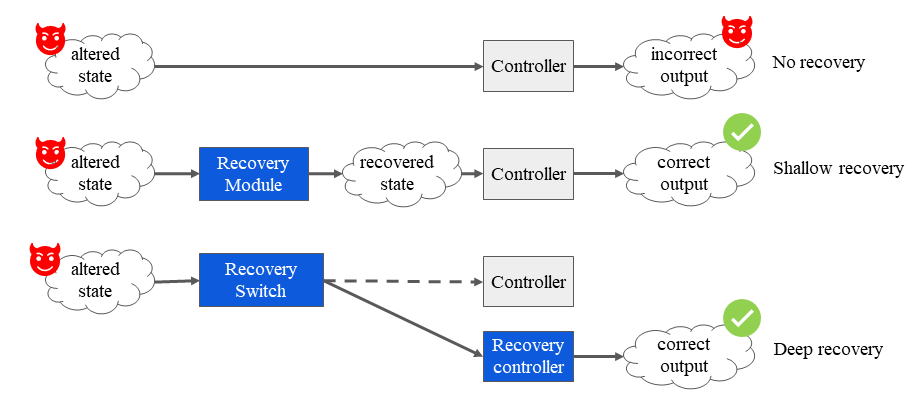}
    \caption{Shallow vs. deep recovery methods.}
    \label{fig:shallow_vs_deep}
\end{figure}

\begin{definition}[Shallow and Deep Recovery]
    \label{def:shallow_and_deep_recovery}
    A shallow recovery is a recovery procedure that does not (necessarily) leverage a controller for recovery purpose only. In contrast, a deep recovery is a recovery procedure that utilizes a dedicated recovery controller.
\end{definition}
For example, in this paper \cite{abad2016restart}, the authors introduce a simplex architecture of controllers, with one of them dedicated for guiding the system to stability to counteract adversarial effects. This method shall fall into deep recovery based on our definition. Another recovery example is to exclude a redundant component, such as a sensor, when the system detects that it is malfunctioning \cite{shin2019rnnDetector}. This method corrects the CPS's physical states without leveraging a dedicated recovery controller, so it is considered shallow recovery based on Definition \ref{def:shallow_and_deep_recovery}. \edit{An illustration of shallow vs. deep recovery is shown in Figure \ref{fig:shallow_vs_deep}. In shallow recovery, the CPS keeps using the same controller, with the corrupted input adjusted by some recovery module. Differently, in deep recovery, the CPS switches to a recovery controller to handle altered behaviors due to attacks.}

In this survey, we review existing research papers that discuss CPS recovery based on Definition \ref{def:recovery}. The papers are sampled from conferences including CDC, RTAS, RTSS, ICCPS and EMSOFT, as well as journals such as IEEE Transactions on Control of Network Systems, IEEE Transactions on Automatic Control and IEEE Transactions on Industry Applications, between year 2014 and 2023. From these venues, we select published works that either propose a concrete recovery solution for a specific scenario, or explore techniques that assist the development of recovery algorithms. Overall, the surveyed publications are tabulated in Section \ref{subsec:list_of_literature}, with a total of \edit{30} papers. This is a relatively small number compared to conventional survey papers, but it indicates that the field of CPS recovery is new and sprouting.

\edit{
\subsection{Comparison to Existing Surveys}
\label{subsec:comparison}
With the increasing integration of computational resources and physical components in cyber-physical systems (CPS), traditional air gaps have been breached, expanding the attack surface. Numerous surveys have identified various attacks and threats in this domain. For instance, some research adapts existing cybersecurity solutions to CPS, presenting an attack taxonomy from a cybersecurity perspective~\cite{kayan2022Cybersecurity}. Recognizing the unique attributes of CPS, surveys, such as Alguliyev et al., ~\cite{alguliyev2018Cyberphysicala}, utilize tree diagrams to illustrate the specific attacks and threats faced by CPS.
The urgent need to fortify CPS has led to a wealth of research focusing on defending against adversarial attacks. A significant portion of this research, as classified by Mitchell et al.\cite{mitchell2014Survey}, revolves around Intrusion Detection Systems (IDS), exploring various detection techniques and audit materials. Giraldo et al.\cite{giraldo2018Survey} extend this by reviewing studies that incorporate physical models in attack detection, acknowledging the intertwining of physical and cyber aspects in CPS.
In light of the complexity of defense strategies, comprehensive reviews such as those by Yaacoub et al.\cite{yaacoub2020Cyberphysical} and Chong et al.\cite{chong2019Tutorial} have emerged, examining multiple facets of defense including attack prevention, detection, and mitigation. Additionally, Olowononi et al.~\cite{olowononi2021Resilient} explore the application of AI and ML in countering cybersecurity threats.
While certain surveys like He et al.\cite{he2016Cyberphysical} focus on specific domains such as smart grids, there's a noticeable tendency for most to emphasize cyber attacks, often overlooking the physical dimension. For instance, Cao et al.\cite{cao2020Survey} concentrate primarily on network attacks within CPS. Although Mahmoud et al.~\cite{mahmoud2019Modelinga} review studies addressing cyber attack issues from a control perspective, but only a fraction of these address control methods in CPS under cyber attacks.
This gap underscores the need for more in-depth reviews that specifically address the recovery of CPS's physical states from attacks to effectively mitigate their impacts.
}

\subsection{Purpose of This Study}
This survey aims to raise scholars' awareness of the burgeoning area of CPS recovery, by reviewing relevant papers in an organized manner and discussing what is currently lacking. Therefore, from this paper, researchers will not only understand the current state-of-the-art, but also identify potential future research directions. \edit{Overall, we have the following contributions.
\begin{enumerate}
    \item We identified the necessity of surveying literature on CPS recovery, due to the increasing impact of CPS and their adversarial attacks. As a response, we surveyed \edit{30} relevant papers from 2014 to 2023 in top venues.
    \item We categorize the surveyed literature from a practical standpoint, i.e., by recovery methods. Researchers can therefore tell what tools have been used by state-of-the-art. In addition, we partition the literature by two major categorizes: methods that do not use control (referred as shallow recovery) and that use control (deep recovery), due to multiple useful tools exist in the control domain.
    \item We analyze the surveyed literature for their detailed techniques, assumptions, and what is being lacked of, to show researchers possible future directions.
\end{enumerate}
}

%% file: sections/2_preliminary.tex
\section{Preliminaries}
\label{sec:preliminaries}

\subsection{Adversarial Attacks on CPS}
\label{subsec:attacks}

With the increasing complexity and openness of CPS, new vulnerabilities are exposed, and different malicious attacks against them are emerging, causing serious personal casualties, social harm, and economic losses~\cite{kong2019state,he2020exploring,kong2018cyber}. For a better presentation of recovery techniques later, this section analyzes attacks targeted on CPS in three dimensions: attack purposes, surfaces, and targets.

The \textbf{attack purposes} fall into two categories according to their impact. The first type of attacks directly undermine the \textit{safety} of CPS. For example, simple disconnections or signal injections into actuators can cause the building or bridge to oscillate at the resonance frequency and cause huge damage to the structures or people in them~\cite{zambrano2021you}.
Because CPSs are usually safety-critical, the following attack dimensions mainly focus on attacks in this category. In addition, the paper focuses more on recovery methods targeted at them.  
The other type of attacks do not directly target the safety of CPSs. Attackers may target the confidentiality of CPSs to prepare for further attack~\cite{chen2021indistinguishability}, the availability of other network infrastructure~\cite{griffioen2020examining}, or people's privacy~\cite{beck2020privaros}.

The \textbf{attack surfaces} belong to the cyber domain or the physical domain.
In the \textit{cyber domain}, attackers can inject false sensor measurements or control inputs, delay or block the signals. For example, spoofing temperature measurements of a device in a power system to smaller values can cause it to overheat and even explode~\cite{farwell2011stuxnet}. In the \textit{physical domain}, attacks can manipulate physical properties to mess up sensor readings~\cite{humayed2017cyber,akowuah2021physical,akowuah2021realtime,quinonez2020usenix,zhang2020detection,yan2020sok,choi2018detecting}. For example, an attacker can use a radio transmitter that broadcasts fake GPS signals to steer a yacht off course~\cite{rutkin2013spoofers}, or use ultrasonic waves to affect IMU sensor readings~\cite{tu2018injected}.  Note that, attacks from the physical domain cannot easily identified using classical software detection mechanisms.

\begin{figure}[!htb]
    \centering
    \includegraphics[width=0.6\textwidth]{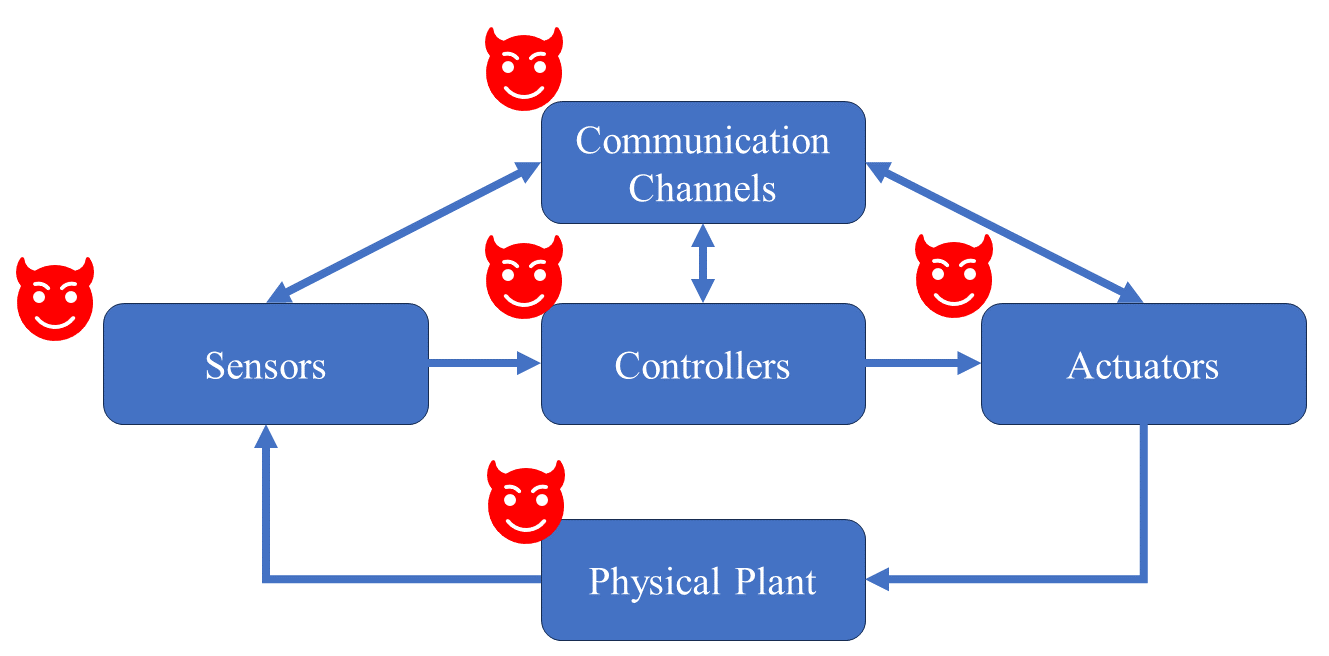}
    \caption{Vulnerabilities of CPS against adversarial attacks.}
    \label{fig:attacks}
\end{figure}

The attacks can \textbf{target} on any components of CPS, thus they are divided into controller, sensor, actuator attacks. \textit{Controller attacks} alter the control logic so that the controller generates incorrect control inputs. One way is to overwrite the control program directly~\cite{albartus2021design}. For example, the Stuxnet can reprogram a Programmable Logic Controller (PLC) to change the frequency of the converters in Iranian Nuclear Facilities~\cite{CARDENAS2012637}. Another way is to change the configurations of the control program. For example, an attacker can compromise the syringe pump via an insecure configuration interface~\cite{surminski2021realswatt}. \edit{Specifically, the command interface of a syringe pump can be hijacked, and a malicious configuration can be sent to the interface to inject a lethal amount of medication to a patient.}
\textit{Sensor attacks} alter the sensor measurements from the cyber or physical domain so that the control logic acts on malicious data that cannot reflect the actual physical states.  For example, the attacker may corrupt LiDAR~\cite{zhu2021can,sun2020robust}, camera~\cite{jing2021too,sato2021dirty,zhou2022doublestar}, GPS~\cite{khan2021m2mon}, and etc. 
There is a trend that attackers tend to avoid being identified by people~\cite{wang2021can}, or compromise multiple sensors simultaneously to bypass defense mechanisms~\cite{cao2021invisible}.
\textit{Actuator attacks} prevent the actuation or control input being implemented by actuators. For example, attackers can block the communication channel to the actuators, or inject false control commands to the actuators~\cite{zambrano2021you}. \edit{Overall, sensors, controllers, actuators and the communication channels in between can all become vulnerabilities of the CPS, as shown in Figure \ref{fig:attacks}.}

\subsection{Detection of Attacks}
\label{subsec:detection}

Although researchers are well aware of the issue of CPS attacks and safety, the current state-of-the-art majorly focuses on attack detection only. So far, detection techniques have well answered the question of how to determine a failure exists, such as EKF and chi-square detector.

Some researchers proposed methods to detect attacks based on system states since the states in CPS are highly correlated due to the system dynamics and physical law. For example, the researchers in \cite{he2020exploring} propose a method to detect attacks based on the correlation between sensor data. Since there are redundancies in the system, it is possible to identify the abnormal data based on the trustful data. It is also possible to detect the attacks in real-time based on the correlation between temporal features \cite{akowuah2021realtime}.

It is also important to notify there are a group of attacks that are hard to be detected or will never be detected. Therefore, it is important to keep the system safe from these attacks. The paper \cite{kim2021stealthy} gives a comprehensive study on the stealthy attacks detection on CPS in real-time. Moreover, the paper \cite{liu2022fail} studies the hidden attacks that can never be detected by the detectors, the optimal hidden attacks are defined to measure the boundary of the negative effect of the attacks. Also, a conservative real-time warning-based framework is provided to guarantee the safety of the CPS even the detectors failed to detect the hidden attacks. \edit{Furthermore, the paper \cite{wang2023attention} utilizes attention mechanisms on top of attack detection to diagnose the attack start time.}

One remark is that, in real-world applications, after detection a CPS must eliminate or alleviate the negative effects caused by attacks, restoring the system back to expected behaviors, and therefore preventing further damages. Therefore, recovery is always required. However, we find that relative few publications have addressed CPS recovery methods, compared to numerous research on detection.

\subsection{Notations}
\label{subsec:notations}
The papers surveyed use different sets of symbols for variables, and we unify their notations as in Table~\ref{tab:notations}, with the same order of appearance in this paper. Moreover, we use subscripts in form of $x_{t}$ to denote a timestamped variable at time $t$. For example, $x_0$ and $x_1$ means the same physical state variable $x$ at time $t=0$ and $t=1$, respectively. In addition, we use superscripts in form of $x^{(i)}$ to denote the same variable corresponding to different system components. For example, $y^{(1)}$ and $y^{(2)}$ mean two measurements of the same variable perceived by sensor 1 and sensor 2, respectively. 
\begin{table}[!htp]
    \centering
    \begin{tabular}{c|c}
        \toprule
        Symbol & Meaning \\
        \midrule
        $x$ & physical state $\in \mathbb{R}^n$\\
        $u$ & control signal $\in \mathbb{R}^m$\\
        $A$ & system dynamics matrices $\in \mathbb{R}^{n \times n}$\\
        $B$ & process dynamics matrix $\in \mathbb{R}^{n \times m}$\\
        $\mathcal{R}$ & stable set of physical states $\subset \mathbb{R}^n$\\
        $w$ & weight of a sensor in sensor fusion $\in \mathbb{R}$\\
        $U$ & control function that outputs $u$\\
        $y$ & measurement $\in \mathbb{R}^p$\\
        $C$ & measurement dynamics matrix $\in \mathbb{R}^{p \times n}$; by default $C$ is identity and $y=x$\\
        $z$ & random noise\\
        $T$ & a period of time $\in \mathbb{Z}_+$ if discrete, $\mathbb{R}_+$ if continuous\\
        $q$ & number of sensors failed $\in \mathbb{Z}_*$\\
        $p$ & total number of sensors $\in \mathbb{Z}_+$\\
        $\hat{x}$ & estimated physical state $\in \mathbb{R}^n$\\
        $\mathcal{X}$ & set of physical states $\subseteq \mathbb{R}^n$\\
        $\mathcal{U}$ & set of controls $\subseteq \mathbb{R}^m$\\
        $J$ & cost function in an optimization problem\\
        $\mathcal{X}_{safe}$ & pre-defined set of safe physical states $\subset \mathcal{X}$\\
        $\mathcal{X}_{target}$ & pre-defined set of target physical states $\subset \mathcal{X}_{safe}$\\
        $r$ & reward in reinforcement learning\\
        $f_s$ & stability cost of a system\\
        $f_c$ & control cost of a system\\
        $\pi$ & policy in reinforcement learning\\
        $\pi^*$ & optimal policy in reinforcement learning\\
        $D$ & set of failed system components in a mass failure\\
        $b$ & utility score of a system component\\
        $v$ & binary variable indicating whether a component is finished recovery in a mass failure\\
    \bottomrule
    \end{tabular}
    \caption{Notations used in this paper in order of appearance.}
    \label{tab:notations}
\end{table}

\subsection{List of Surveyed Literature}
\label{subsec:list_of_literature}

\begin{figure}[!htb]
    \centering
    \includegraphics[width=\textwidth]{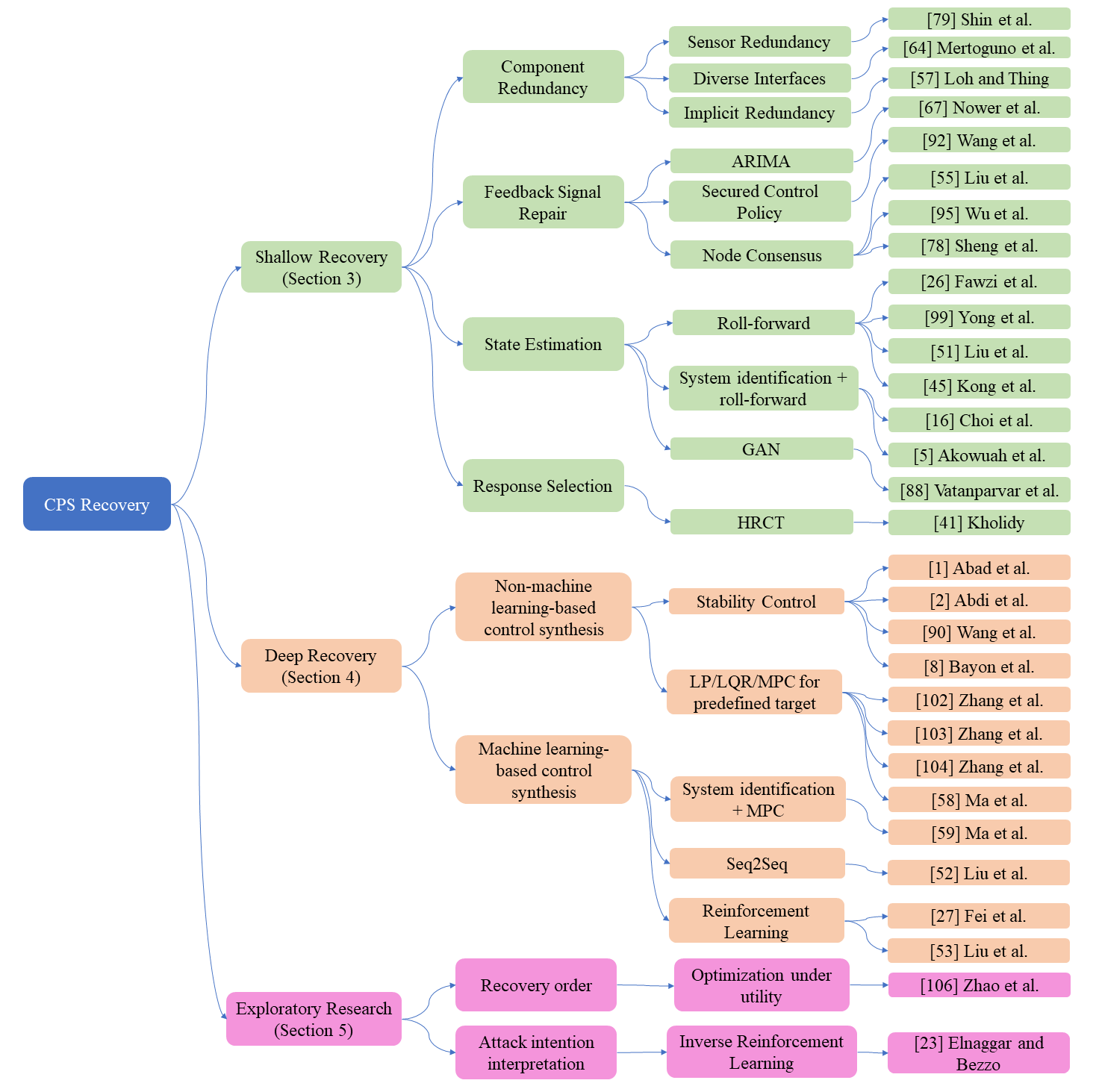}
    \caption{Tree diagram of the surveyed literature.}
    \label{fig:tree}
\end{figure}

Based on Definition \ref{def:shallow_and_deep_recovery}, we identify 16 papers on shallow recovery and 11 papers on deep recovery.
Moreover, we identify two papers \cite{zhao2016interdependent,bezzo2018irl} that do not purpose recovery solutions, but aim to discover information that assist recovery instead. We believe these papers are also worth mentioning in this survey, and include them in Section \ref{sec:exploratory}. Finally, we provide a discussion in Section \ref{sec:discussion}. \edit{The 16+12+2=30 papers are categorized as illustrated in the tree diagram of Figure \ref{fig:tree}.}

%% file: sections/3_shallow.tex
\section{Shallow Recovery}
\label{sec:shallow}
In this section, we discuss literature on \textbf{shallow recovery} techniques. By Definition \ref{def:shallow_and_deep_recovery}, these are recovery solutions that do not leverage dedicated controllers. Researchers have proposed various methods to guide a CPS back to its desirable conditions without solving for control.

\subsection{Component Redundancy}
\label{subsec:redundancy}

One simple recovery method is to leverage component redundancy, so that we can perform operations such as excluding suspicious components. Here, redundancy means that multiple components serve for the same purpose and it is acceptable to discard some of them while keeping the system running. \edit{The procedure that excludes corrupted components is referred to by researchers as detection, isolation and reconfiguration (DIR) \cite{cpsKnowledgeArea2019}, i.e., detecting suspicious components, isolating them from participating in the process, and optionally reconfigure other components. If no reconfiguration is done, this procedure can also be denoted as DI (without R) only.}

\begin{figure} [!htb]
    \centering
    \includegraphics[width=0.7\textwidth]{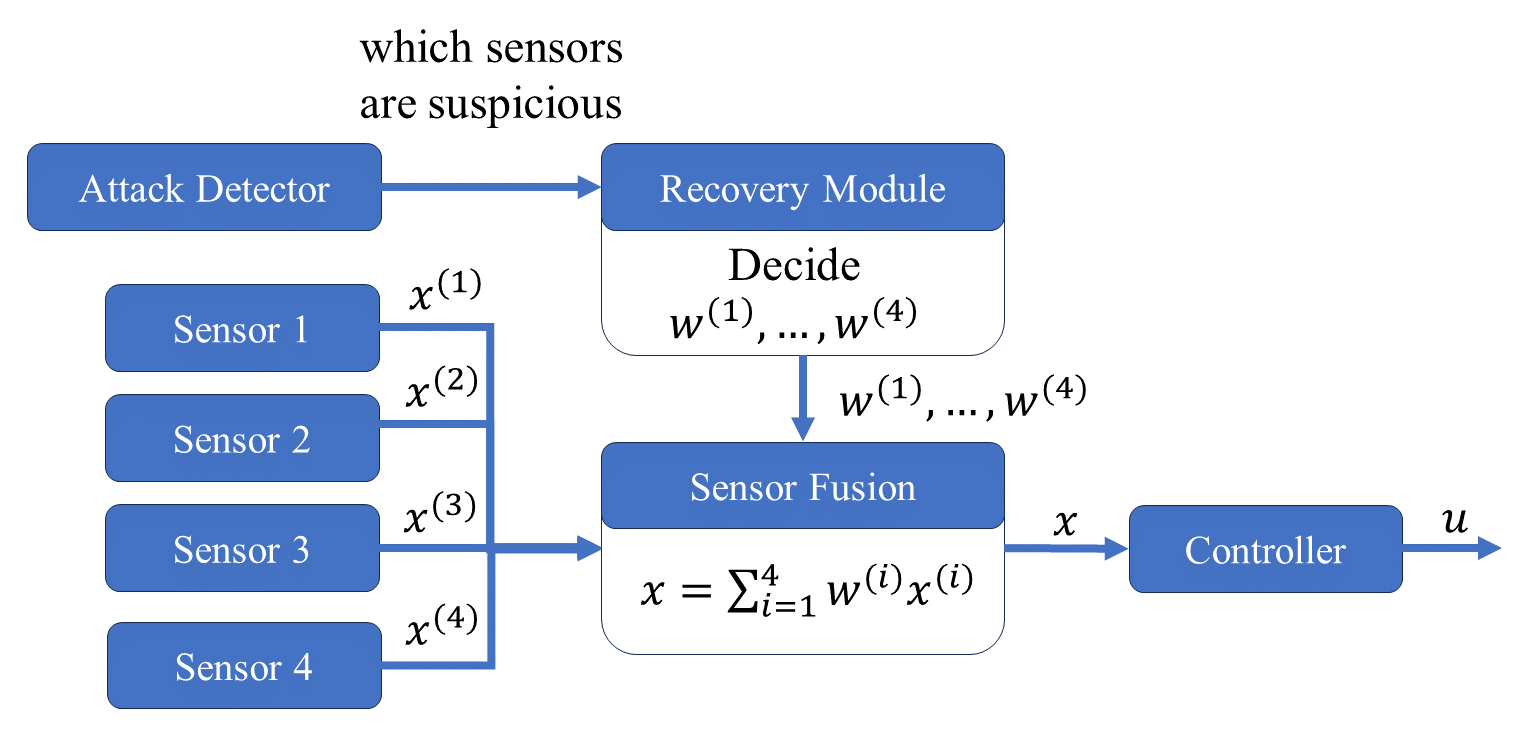}
    \caption{Workflow of component exclusion method discussed in \cite{shin2019rnnDetector}. The recovery module assigns weights to redundant sensors based on how much each of them can be trusted. The assumption is multiple sensors are serving for the same function, i.e., redundancy.}
    \Description{Workflow of component exclusion of four redundant sensors, with the recovery module assigning weights to the four sensors for fusion.}
    \label{fig:component_exclusion}
\end{figure}

Based on the description above, DI or DIR only needs a powerful detector with high sensitivity and acceptable precision to inform the system on degree of trust/suspicion. For instance, researchers have shown that recurrent neural networks (RNNs), a deep learning model that captures pattern in time series \cite{medsker2001recurrent}, can be trained as a detector and raise alarms on a redundant set of system components, while telling how suspicious each individual sensor is \cite{shin2019rnnDetector}. This work demonstrates an elementary scenario: for a car travelling on a straight road, the speed of its four wheels, monitored by four distinct sensors, should be approximately the same. If the four sensor readings are different, consider an attack and assign a smaller weight to the suspicious wheel(s) when doing weighted-average sensor fusion. Notice that the component exclusion/isolation can be either hard or soft, such that a hard exclusion completely bans the component from participating any further system operation, e.g., assigning a zero weight to a suspicious wheel speed sensor. On the other hand, a soft exclusion only mitigates its participation, e.g., assigning a non-zero but smaller weight. The workflow is illustrated in Figure \ref{fig:component_exclusion}. 

Similarly, another study has proposed a Byzantine fault-tolerant architecture, namely BFT++, by leveraging redundant component replicas with diverse software interfaces \cite{mertoguno2019physics}. Based on traditional CPS designs that are resilient to Byzantine faults, if a sufficiently large portion of redundant components remains intact within a small window of vulnerability, the system will recover from adversaries. The assumption is that all the redundant components consume the same input, and the attacker can only interrupt one input signal within a period of time. Therefore, implementing diverse interfaces across all the redundant components may lead to a higher chance for tolerance and recovery. In the design of BFT++, input signals enters an artificial divert stage before entering each component. One example diversification method is to customize each component's executable memory layout, such that the same variable does not live at the same address for all components. Another example is to utilize different instruction set architectures (ISA) on the components, such as using x86 on one and ARM on another. This diversity can be further encouraged if the input into these redundant components must undergo different delays, creating temporal differences.

Another research work focuses on exploiting implicit redundancy \cite{loh2023enhancing}. While explicit redundancy means identical physical components serving for the same functionality, implicit redundancy means non-identical components mimicking each other's functionalities by additional software processes. For example, a vehicle's camera, radar and LiDAR take different forms of inputs, but we can interpret the same output, such as the distance to an obstacle ahead, from all three components. The advantage of implicit over explicit redundancy appears when the duplicate physical components are expensive. Hence, when a subset of components are failing due to adversaries, it is possible to substitute its functionality using another. The authors of this paper provide a search algorithm for substitutions on the graph of the CPS information flow. This algorithm is proved to preserve desirable properties, from a predecessor node in the graph to a successor, when the latter mimics the former. For example, if a camera has a certain guarantee on identifying obstacle distances, then a LiDAR mimicking the camera's output shall inherit the same guarantee.

\subsection{Feedback Signal Restoration}
\label{subsec:feedback_repair}


When a measured signal is lost or losing integrity due to adversarial attacks, a type of recovery methods aim to restore a reliable signal, and therefore guide the CPS back to target physical states. 

One scenario is that a signal is partially corrupted, and we are able to use the healthy readings to estimate the true value of the corrupted ones. A corresponding technique that utilizes autoregressive integrated moving average (ARIMA) is introduced in~\cite{lim2014arima}. ARIMA is a machine learning method that forecasts a time series, such that a forecasted data point fits a regression pattern along with other data points \cite{ho1998arima}. Therefore, it is useful to repair missing sensor readings or fix abnormal sensor readings during CPS runtime, under the assumption that a healthy sensor produces a smooth trend of data. Specifically, the reason for adopting ARIMA model is because systems have inertia, so that any sensor measurement must not change too drastically and approximately maintains a linear relationship with its auto-regressive past values. The recovery approach built on top of ARIMA, namely Efficient Temporal and Spatial Data Recovery (ETSDR), trains an ARIMA model from pre-collected data offline, and then uses this model as a substitute of system dynamics. Next, at runtime, when a sensor reading has large difference, either temporal (very different from the sensor's own past readings) or spatial (very different from neighboring sensors), the recovery module replaces the reading with ARIMA prediction plus a small drift value. The ARIMA model itself is also constantly updated by confident sensor readings or predicted state values during runtime. 

The authors of \cite{lim2014arima} demonstrate the capability of ETSDR to repair missing data by comparing it to other time series forecasting methods for CPS sensor readings, including weighted prediction (WP) \cite{xia2011wp} and exponentially weighted moving average (EWMA) \cite{choi2012ewma}. They generate simulated sequences of sensor reading data and perform stochastic data loss from 10\% to 60\%. Then, three error metrics, root mean square deviation (RSME), mean absolute error (MAE) and integrated absolute error (IAE), are used to measure the difference between model-predicted data and actual data. 



A more specific scenario is sensor networks under denial of service (DoS) attacks \cite{wang2022security}. Here, the CPS's components may utilize a network system, such as the Internet, to communicate with each other. The authors use a networked inverted pendulum system as an example, with the sensors and actuators reading from and acting on the pendulum, but the controller being remote. DoS attacks aim to cause packet losses on the feedback channel to the remote controller. Leveraging game theory, the authors propose an augmented control policy to handle the losses. This policy selects an optimal control output that takes advantages of available packet information to maximize the CPS's utility.

DoS attacks are also discussed for microgrid CPS recovery strategies \cite{liu2022consensus}. Here, a distributed controller remotely talks to a cluster of information nodes and another cluster of physical power grid nodes. The controller has three objectives: (1) synchronizing power frequency among grid nodes, (2) reducing error of power frequency based on a reference frequency, and (3) maintaining consistent incremental costs on all grid nodes. Upon a DoS attack, part of the information required to control the grid nodes is lost. As a countermeasure, this paper proposes a two-step consensus algorithm to repair the partially lost signal. First, multiple intrinsic mode functions (IMFs) are obtained from historical data via statistical inferences, specifically empirical mode decomposition (EMD). These functions do not necessarily have explicit meanings, and will reach a consensus to repair every lost value. The consensus mechanism is carried out by learning a deep neural network in the form of extreme learning machine (ELM), with IMF estimations as inputs and repaired signal values as outputs.

The above paper leverages a deep learning technique for the final repair decision, and the following work proposes an alternative deep method using a modified denoising auto-encoder (MDAE) \cite{wu2022integrated}. That is, the CPS leverages two deep neural networks that serve as encoder and decoder respectively before and after the network communication. The partially corrupted signal enters the encoder and become encrypted, and it is the decoder's job to not only decipher the encrypted signal, but also repair the signal while decoding. Ideally, this technique can both counteract the effects of partially corrupted signals and defend the CPS against eavesdropping.

Similar to \cite{liu2022consensus}, a succeeding work \cite{sheng2023collaborative} also introduces a consensus method to recover from failed CPS networks. Here, nodes are assumed to be connected via a communication network abstracted as a graph. A loss of connection between two nodes is therefore equivalent to discarding an edge and modifying the graph's topology. The corresponding recovery strategy is based on a collaborative shortest-path algorithm, namely dynamic routing. Specifically, the algorithm iteratively searches for a broken edge that can restore the connection to the largest number of nodes if repaired. By computing this order, the links can be gradually rebooted for signal repair.

\subsection{State Estimation}
\label{subsec:state_estimation}

Another adopted recovery method is to predict or estimate the actual physical state that an unattacked system should be in. \edit{Different from feedback signal repair, which restores lost measurements using reliable measurements in proximity, this approach usually aims to handle failures in state sensor integrity, with measurements no longer reliable.} One remark is that the estimated physical state can also assist a dedicated recovery controller in deep recovery, which will later be discussed in Section \ref{sec:deep}. Here, we only cover relevant shallow recovery methods. Another remark is that the synchronization between cyber states and physical states has been a widely studied topics, but to our knowledge, only a few papers (discussed below) have adapted this method to conquer problems in CPS recovery.

The most straightforward state estimation is to compute states in a model-predictive way, i.e., roll-forwarding, assuming that the dynamics is a known white-box and has negligible noises. The earliest paper \cite{fawzi2014secure} on this topic was published in 2014 and assumes a noisy linear dynamics as
\begin{equation}
\label{eq:fawzi_dynamics}
    \begin{split}
    x_{t+1} &= Ax_t + B\big(U_t \big(y_0, \dots, y_t \big) + z^{(1)}_t \big)\\
    y_t &= Cx_t + z^{(2)}_t
    \end{split}
\end{equation}
where the notations are defined in Table~\ref{tab:notations}. The estimation problem is how to correctly reconstruct the initial state $x_0$ after a period of time $T$, when the sensor readings $y_0, \dots, y_{T-1}$ are available but they are corrupted by attacks on a fixed number of $q$ dimensions, i.e., $q$ sensors fail and give erroneous measurements during runtime. The authors then define correctable errors, which can be tolerated when reconstructing $x_0$. Based on this idea, the major takeaway is that the maximal size of $q$ is $\lceil p/2 - 1 \rceil$ such that the error is correctable, where $p$ is the total number of sensors. In plain words, with at most half of the sensors failed, we can still correctly estimate $x_0$ using the authors' algorithm, and therefore predicting the succeeding states from the dynamics. Consequently, the system can constantly recover itself by iteratively running the estimator and replace the faulty sensor readings with estimated values when producing control. One drawback of this approach is that the computation is expensive when the number of corrupted sensors is close to the upper-bound $\lceil p/2 - 1 \rceil$.
Also, the approach is unable to handle the case that more than half sensors are corrupted.

\begin{figure}
    \centering
    \includegraphics[width=0.6\textwidth]{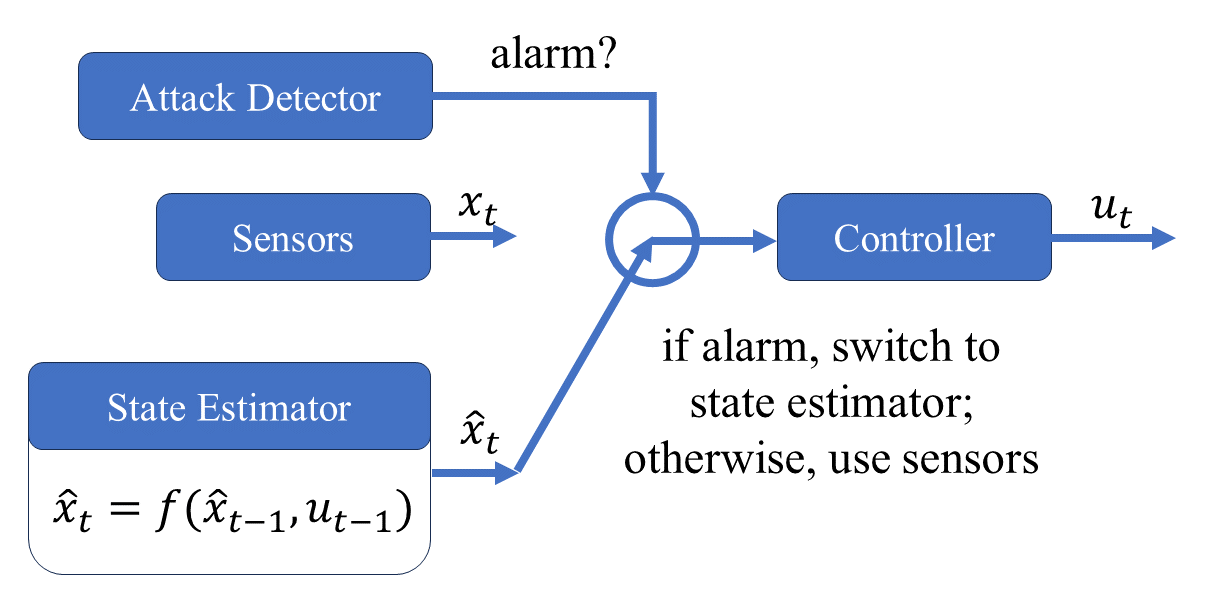}
    \caption{A typical state estimation-based shallow recovery. Upon a sensor detection, the system switches from using physical sensor readings to estimated states. A checkpointer \cite{kong2018checkpointing} can be optionally used to accelerate state estimator by caching information.}
    \Description{Workflow of state estimation-based recovery, with the detector deciding to use either the physical sensor readings or estimated states as the input into controller.}
    \label{fig:shallow_roll_forward}
\end{figure}

Different from the paper in the above paragraph, \cite{kong2018checkpointing} considers a more relaxed assumption, where the sensor corruption does not happen at the very beginning, but rather some time in the middle. In this case, the computation will be much cheaper, with a checkpointer being leveraged to memorize a sliding window of sensor readings. The fixed-size sliding window, or detection window, enqueues a new measurement and its corresponding control signal upon reading and dequeues the oldest historical data at the same time. Under the assumption that the checkpointer remains uncorrupted, the system validates the consistency within the detection window and reports anything suspicious. Then, the same recovery method applies: replacing the flawed sensor reading with the estimated state from dynamics roll-forwarding. The workflow of the recovery procedure is illustrated in Figure \ref{fig:shallow_roll_forward}. This tool is compared to later works in experiments and please see Section \ref{subsec:white_box} for details.

Although papers on state estimation mostly discuss corrupted sensors caused by general attacks, some specify the type of attack and design corresponding solutions. For example, the paper \cite{yong2015resilient} particularly considers switching attacks. By its name, switching attacks, or switching location attacks, are a type of attacks where the adversary constantly changes their attack location at some frequency. For example, sensor \#1 is under false data injection from $t=1$ to $t=3$ seconds, and then the attack stops and changes to sensor \#2 from $t=3$ to $t=5$. This is opposite to fixed location attacks, where the adversary stays still. A typical target for switching attacks is on interconnected distributed systems, such as a multi-area power grid, with each area a controller. The attacker aims to break the connection topology by either shutting down a node (a sensor or an actuator), forcing it to go offline, or interrupting an edge to cut a link between nodes. Under this type of failure, the objective of this paper is to perform an unbiased estimation of the actual system state, including all components, without much knowledge of the attack except for an upper bound of the number of failed sensors and actuators. To perform the state estimation, a noisy (linear) CPS under switching attacks is modeled as a (linear) stochastic system with hidden discrete modes, with each mode includes the information of the CPS's operation modes, current topology, and if the sensors/actuators are under attacks. Next, mode-matching is done by computing the probability of which mode the system is in, given the available information. The paper also provides a theoretical analysis on the method's resilience, such as the sufficient conditions for the estimated stochastic model to represent the true underlying system.

Another paper \cite{liu2016dynamic} also proposes a method to estimate states under switching location attacks. Nevertheless, this paper discusses a problem of restoring an unknown initial state $x_0$ based on sensor measurements from $t=0$ to some $t=T$, while some sensors are under switching attacks. In other words, a subset of sensors are giving erroneous readings, while this subset is changing at some constant frequency. The paper hence provide a solution, namely a dynamic decoder, which is a function that maps measurements $y_0, \dots, y_T$ to an estimated initial state $\hat{x}_0$. The approach to solve for this decoder is by convex optimization, searching for $\hat{x}_0$ that minimizes $\sum_{t}||y_t - CA_t\hat{x}_0||_1$, where $C$ is the linear mapping matrix from states to measurements, and $A_t$ is the roll-forwarding matrix from the initial state to the state at $t$, based on system dynamics. The paper proves the sufficiency and necessity of this optimization problem to compute an accurate decoder function.

When dynamics is an unknown black-box, state prediction becomes a more complicated problem. The easiest way is to break the procedure into two steps: (1) estimate the dynamics itself with a surrogate model and (2) call a state estimation algorithm on the surrogate model. System identification techniques have paved the way for the first step. For example, observer/Kalman filter identification (OKID) \cite{kalmanParams1993, kalmanfilter2003} has been in the game since 1990s. As the field of machine learning matures, scientists carry on system dynamics identification with more up-to-date models, from recursive least square (RLS) and least mean square (LMS) \cite{recursiveLeastSquare2004} to deep learning for complex and even chaotic dynamics \cite{jiahao2021chaos}. Approaches other than this two-step scheme exist, but they are mostly associated to deep recovery and will be discussed in Section~\ref{sec:deep}.

System identification can be accomplished by simple statistics in matured toolboxes when there is sufficient prior knowledge. For instance, the paper \cite{choi2020software} assumes a template for a linear quadcopter dynamics which consists of 12 dimensions including position, pitch angle, linear and angular velocity. Another assumption is the full knowledge of a PID controller. Then, with the Matlab system identification (SI) toolbox \cite{ljung2012matlab}, the linear dynamics can be estimated by iterative regression on a set of pre-collected training data on system inputs and outputs. Then, recovery is accomplished by replacing the faulty sensor measurement with the estimated states by the identified surrogate system. 


Deep learning models, i.e., neural networks, have also been used to capture system dynamics. The motivation for utilizing deep models is that they are designed to discover complex patterns with implicit influencing factors, which can be useful for modeling unknown real-world system dynamics on one hand. On the other hand, a typical disadvantage of deep models is that these models require large amount of data to train or will end up overfitting \cite{li2019overfitting}. In \cite{kong2021learning}, the surrogate model to capture the system dynamics is an LSTNet \cite{lai2018lstnet}, a neural network architecture that is designed to perform long and short-term time series forecasting tasks. The architecture contains convolutional layers \cite{o2015convolution}, which are designed to find location-agnostic patterns, e.g., time-invariant/context-invariant system dynamics, as well as recurrent layers \cite{medsker2001recurrent}, which aim to capture temporal dependencies and can be useful in fitting closed-loop dynamics. However, like other deep models, LSTNet requires a pre-training phase on data collected in a normally behaved system, and has a chance of overfitting due to its complex nature. The recovery in this paper is simple: replacing the sensor reading with LSTNet estimation after an attack is detected. Moreover, the paper adopts the checkpointer in \cite{kong2018checkpointing} for caching runtime data points.

\begin{figure}
    \centering
    \includegraphics[width=0.75\textwidth]{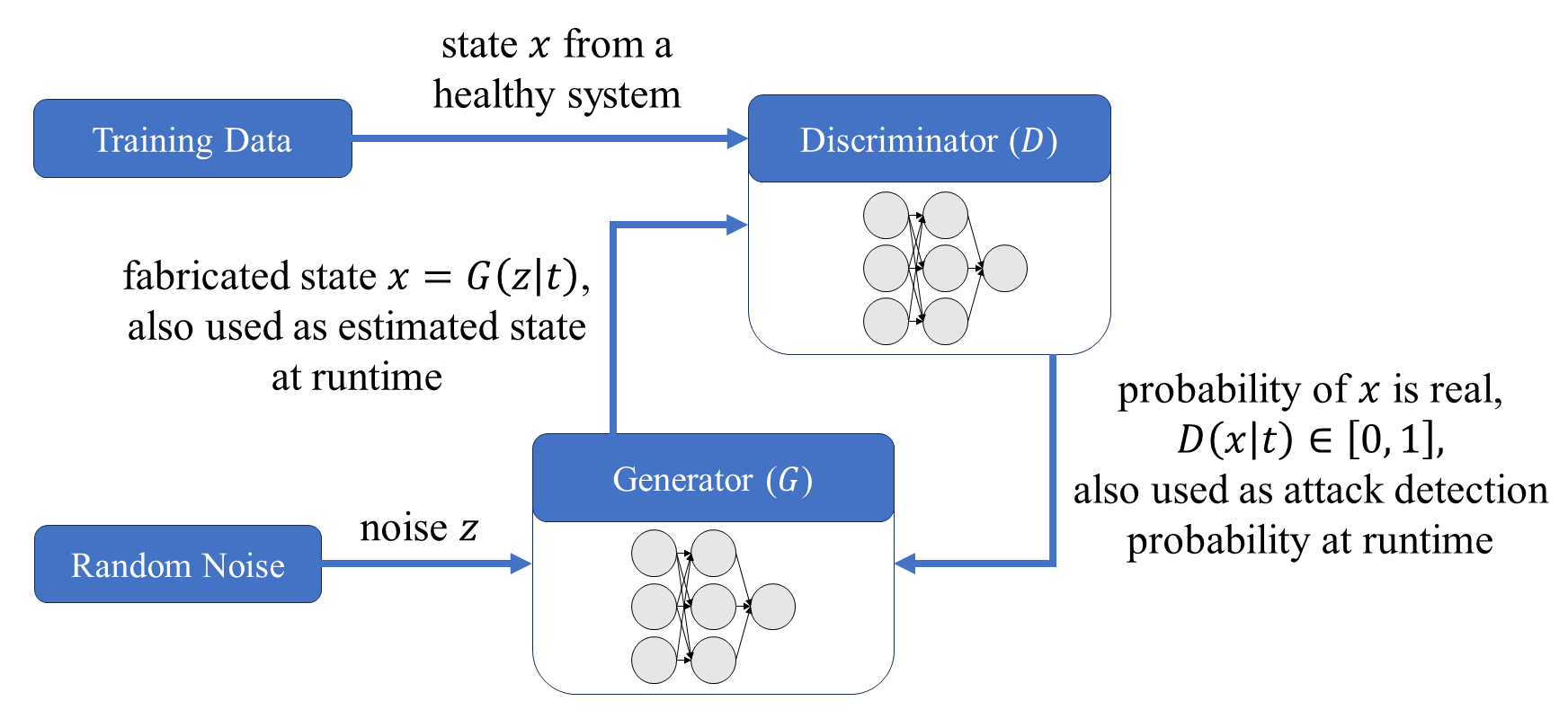}
    \caption{The training procedure of a CGAN discussed in \cite{alfaruque2019gan}. At runtime, G and D will be used respectively as the state estimator and detector as in Figure \ref{fig:shallow_roll_forward}.
    }
    \Description{Workflow of training a CGAN: the discriminator D and the generator G iteratively exchanges information. }
    \label{fig:cgan}
\end{figure}

Another state estimation method is conducted by generative adversarial networks (GAN), described in \cite{alfaruque2019gan}. The model of GAN \cite{creswell2018generative} has been proved useful for generating data points following an unknown distribution. It consists of a discriminator neural network (D) that is trained to distinguish out-of-distribution data (fake/fabricated, output=1) from in-distribution data (real, output=0), as well as a generator neural network (G) trained to produce fake data from random noise input to fool D. These two neural networks co-evolve through competition and eventually G can generate data points almost indistinguishable from the real ones. 
When learning an underlying dynamic that produces CPS physical states, different periods of time can be seen as different underlying distributions where states are drawn from. Therefore, time is a condition and the GAN is in fact a conditional GAN (CGAN). A major contribution of this paper is how to utilize D and G after their convergence. First, the authors claim that D can serve as an anomaly detector to capture faulty states, because a well-trained discriminator should able to distinguish whether a state is from the learned dynamics or not, given the temporal condition. On the other hand, G can serve as a state estimator. The idea is that the data points generated by a well-trained generator should be close to true data distribution, i.e., healthy system dynamics. The training process (not runtime) of this approach is illustrated as in Figure \ref{fig:cgan}. At runtime, the recovery procedure is still as in Figure \ref{fig:shallow_roll_forward}, with the detector replaced by trained D and the state estimator replaced by trained G.

The authors of the CGAN paper \cite{alfaruque2019gan} also evaluates their approach with experiments. On a battery management system of an autonomous vehicle, two types of attacks are simulated: (1) a physical attack that changes the underlying dynamical system, such as replacing the battery with an alternative, low-performance battery, and (2) a denial-of-service (DoS) attack, where all sensor readings are injected with a random noise. 

\subsection{Response Selection from a Candidate Set}
\label{subsec:response_selection}

Finally, researchers have proposed wrapper tools around available recovery methods. In one particular paper \cite{kholidy2021autonomous}, the assumption is that the system has a set of candidate responses to an attack, and the objective is to select the one with maximized benefit and/or minimized cost. Notice that these responses can be either shallow or deep, but this research only considers the ones without using a dedicated recovery controller, i.e. shallow techniques. Therefore, we attribute it to shallow, but notice that this framework can be trivially extended to deep recovery.

The paper particularly discusses this approach and designs an autonomous response controller (ARC). One prerequisite for this controller is a hierarchical risk correlation tree (HRCT), which is a stochastic decision tree, with each path being a path an attacker can take to reach certain malicious goals. For example, an attacker can first access the authentication of the system's cached key, then modify the key, then inject false data to sensors as an imposter, and finally accomplishes a man-in-middle attack. This tree structure also includes a quantitative risk, or cost, at each path. The other prerequisite for ARC is a candidate set of responses, which is application-specific. The authors separate these responses into two major categories: (1) cyber countermeasure responses, which resolves problems by calling a software process, such as packet dropping, traffic isolation, and process termination, and (2) physical countermeasure responses, which requires changes in physical components, such as deploying a secure gateway. The job of the ARC is to select a response, or a subset of responses, to mitigate the cost based on HRCT. At runtime, nodes of HRCT are partially observable, and the probability of every path taken can be computed. With every response associated with a quantitative impact on the tree nodes, such as traffic isolation can reduce the probability of false data injection, minimizing the total cost can be achieved by solving an optimization problem.

\begin{figure}[!htb]
    \centering
    \includegraphics[width=0.65\textwidth]{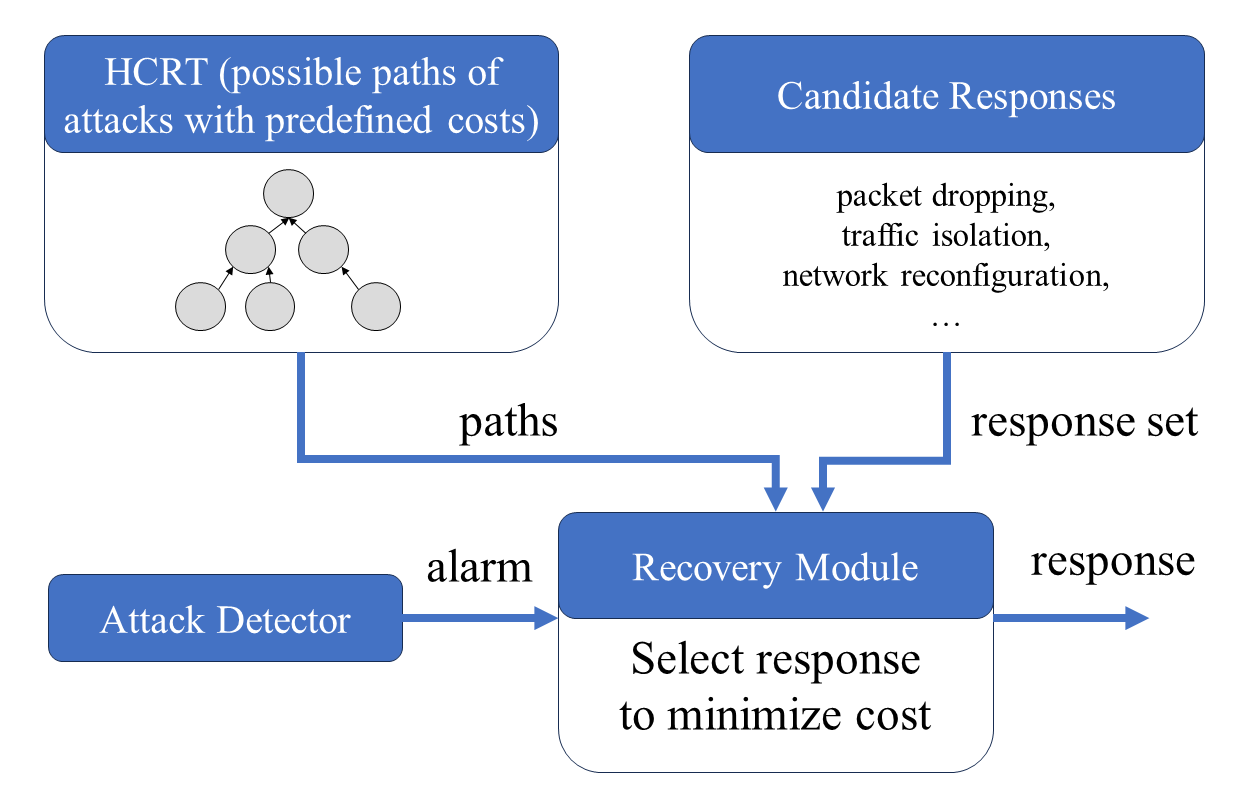}
    \caption{The workflow of response selection as discussed in \cite{kholidy2021autonomous}.}
    \label{fig:response_selection}
    \Description{Workflow of response selection, with the recovery module query information from an HRCT and a candidate set of responses. It then outputs a selected response.}
\end{figure}

%% file: sections/4_deep.tex
\section{Deep Recovery}
\label{sec:deep}
This section reviews papers on deep recovery, which means a dedicated controller is used for guiding the CPS back to its desirable conditions. Therefore, deep recovery means that the recovery problem is boiled down to a control problem. \edit{These control synthesis techniques are further divided into non-machine learning-based and machine learning-based synthesis.}


\edit{\subsection{Non-Machine Learning-based Control Synthesis}}
\label{subsec:white_box}
When the dynamics is known, the recovery controller is able to be solved by conventional control methods, including linear programming (LP), linear quadratic regulator (LQR), model predictive control (MPC). We identify two control targets from the corresponding literature: system stability and system safety.

Existing papers have discussed system designs and algorithms to reach a stable set of physical states. For instance, \cite{abad2016restart} introduces the simplex architecture \cite{sha2001simplex}, which means the system is divided into two isolated parts: a main unit with a normal, unverified and complex controller serving for the CPS's task purposes and a rescue unit with a verified and simple controller, with a sole aim to stabilize the system. The assumption is that only the main unit is vulnerable, and the system dynamics is (or can be approximated as) a linear time-invariant system, in form of $x' = Ax + Bu$, where $x$ is the state and $u$ is the control signal. Moreover, the control function is also linear, in form of $u = Kx$. Under such linearity, there exists some coefficient matrix $P$ that defines a stable ellipsoid of states $\mathcal{R} = \{x \mid x^TPx < 1\}$, where the Lyapunov potential function $x^TPx$ has a negative-definite derivative. Therefore, any trajectory starts within $\mathcal{R}$ never leaves $\mathcal{R}$, and the rescue unit's objective is to guide the system's physical state back to $\mathcal{R}$ and keep it inside for an unbounded time. When the main unit fails due to attacks, the rescue controller takes over the system and ensures its stability, while the main unit being rebooted. Since only the main unit of the complex controller is assumed to be vulnerable, the sensors are still intact and sensor readings can still be used by the simple controller dedicated for recovery.

\begin{figure}
    \centering
    \includegraphics[width=0.55\textwidth]{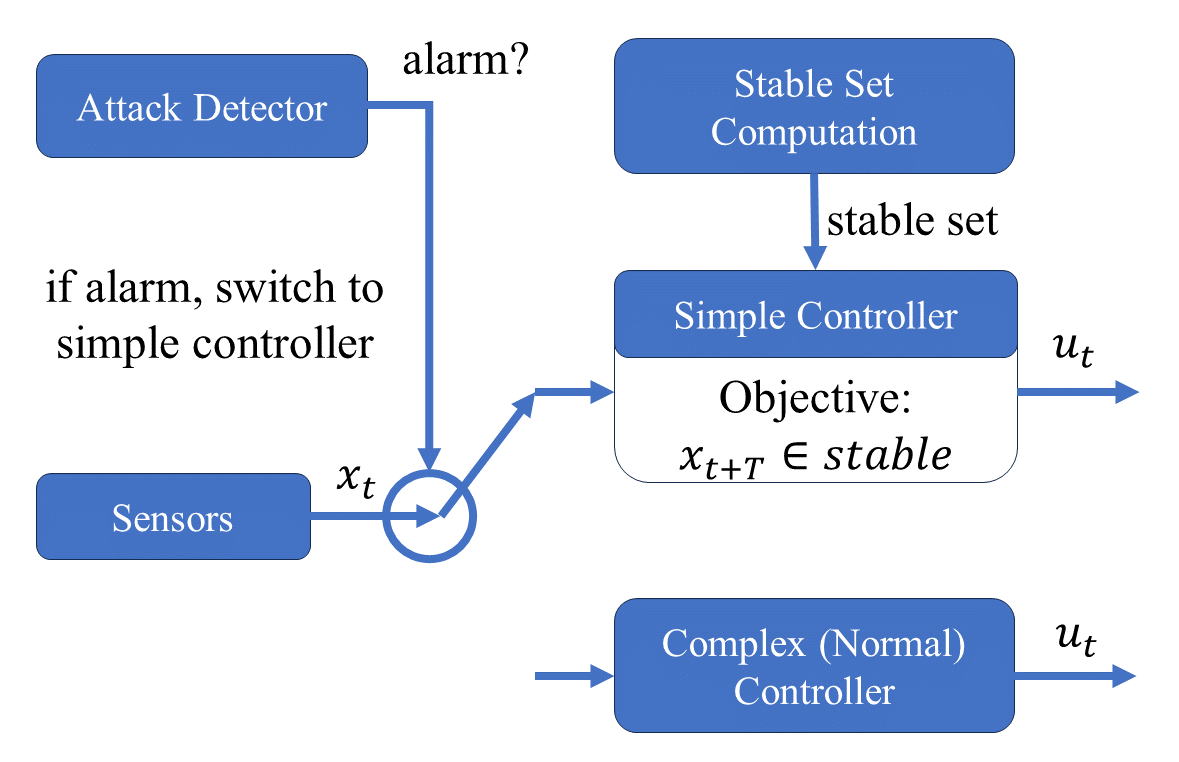}
    \caption{Restarting recovery workflow on simplex architecture in one iteration, proposed by \cite{abad2016restart, abdi2018restart}. Upon an attack detection, the system switches from complex to simple controller, while constantly checking if the complex controller can be rebooted.}
    \Description{Workflow of restart-based recovery on simplex architecture as proposed in \cite{abad2016restart, abdi2018restart}.}
    \label{fig:restart_simplex}
\end{figure}

A succeeding work \cite{abdi2018restart} enhances the stability guarantee of this simplex architecture approach by finding secure execution intervals (SEI). In this work, the authors address a more severe scenario where an adversary is intentionally driving the system state outside of the stable ellipsoid. Therefore, SEI is defined as a time interval that the system is guaranteed to stay inside the stable set despite an adversary. When a system starts inside the stable ellipsoid, due to system inertia, the unstable states are not immediately reachable but rather requires time. A period of SEI hence exists no matter how hard the adversary pushes the system. The recovery module sorts the components to be rebooted into multiple tasks in the order of priority while searching for the next best restarting time within SEIs using reachability analysis. Each rebooting task is then assigned to a sufficiently long SEI in which the task is guaranteed to finish. Consequently, this paper presents a restarting-based CPS recovery method that guarantees to counteract an unstablizing adversary. 




Specific CPS applications have leveraged stability-guided recovery controllers. For example, when the dynamics is known, legged robots can perform a black-flip to achieve stability in standing-up positions \cite{wang2022kinetostatic}. Aiming for the stable physical states, quadruped robots are able to leverage recovery-dedicated controllers to perform corresponding foothold motions after falling. Another application considers exoskeleton in healthcare industry \cite{bayon2022cooperative}. In this paper, the goal is to stabilize a pair of exoskeletons around a person's ankles while an adversarial pushing force is exerted on the person's body. After detecting the perturbing force, the exoskeletons must react to maintain not only a stable position but also stable forward steps. This is as well guided by stability-guided control.

\begin{figure}
    \centering
    \includegraphics[width=0.8\textwidth]{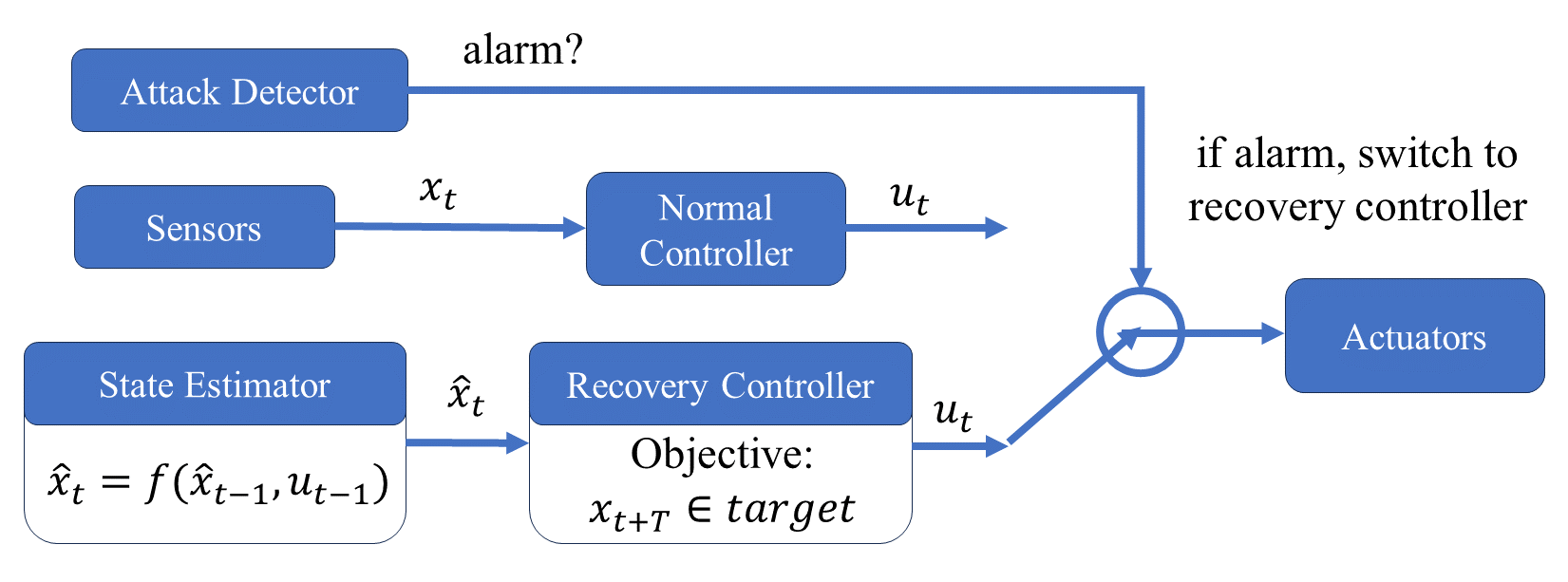}
    \caption{State estimation-based deep recovery, used in \cite{zhang2020lpRecovery,zhang2021lqrRecovery,ma2020rmpc,ma2021drmpc}. Different from Figure \ref{fig:shallow_roll_forward}, upon a detection, the system not only uses estimated states to replace physical sensor readings, but also switches to a recovery controller to drive the system back to a pre-defined target state set. A checkpointer \cite{kong2018checkpointing} can be optionally used to accelerate state estimator by caching information.}
    \Description{Workflow of state estimation-based deep recovery, where the detector decides a path from two: (1) physical sensors to normal controller, or (2) estimated states to recovery controller.}
    \label{fig:deep_roll_forward}
\end{figure}

Different from stability, another control target is for the CPS's safety, which is usually defined as reaching a given subset of the physical state space, i.e., $\mathcal{X}_{safe} \subset \mathcal{X}$. The authors of \cite{zhang2020lpRecovery} continue to leverage the state prediction method from \cite{kong2018checkpointing} in a deep recovery context. Here, a recovery controller takes over the system upon a failure is detected, and computes control signals $u_0$ to $u_T$ in a time window $T$ by solving the linear programming (LP) problem as in~\eqref{eq:lp_recovery}, assuming time is discrete.
\begin{equation}
    \label{eq:lp_recovery}
    \begin{split}
    &\text{minimize}_{x_1, \dots, x_T \in \mathcal{X},\, u_0, \dots, u_T \in \mathcal{U}}\, J(x_1, \dots, x_T, u_1, \dots, u_T)\\
    &\text{subject to } (x_0 = x_{init})\, \land
    \bigwedge_{t=0}^T (x_{t+1} = Ax_{t} + Bu_{t})\, \land 
    \bigwedge_{t=0}^T (x_t \in \mathcal{X}_{safe})\, \land 
    (x_T \in \mathcal{X}_{target})
    \end{split}
\end{equation}
Here, the normal behavior is defined by the system states being safe and inside a target set, i.e., $x \in \mathcal{X}_{safe}$ and $x \in \mathcal{X}_{target}$. The constraint requires (1) the first state is a given initial state $x_{init}$, (2) the system follows a given dynamics, and (3) eventually the system will be guided back to normal at time $T$. The second constraint utilizes the roll-forwarding state prediction technique in \cite{kong2018checkpointing}. Moreover, the cost function $J$ is linear in this case, and a succeeding work \cite{zhang2021lqrRecovery} replaces it with a quadratic cost function such that the recovery controller becomes a linear quadratic regulator (LQR), claiming that LQR recovery provides a smoother state trajectory. Optionally, there can be an additional constraint $x_T, \dots, x_{T+M} \in \mathcal{X}_{target}$, which asks the controller to not only bring the system back to the target, but also maintains it there for a period of length $M$. The workflow of this recovery procedure is illustrated in Figure \ref{fig:deep_roll_forward}. Another succeeding work \cite{zhang2023real} leverages model predictive control (MPC) on a more relaxed constraint on system dynamics. Here, the dynamics is allowed to be nonlinear, and the recovery control utilizes a Taylor polynomial approximation to simplify the computation.

The authors of \cite{zhang2021lqrRecovery} compares the shallow recovery in \cite{kong2018checkpointing}, linear-cost deep recovery in \cite{zhang2020lpRecovery} and their own quadratic-cost method. Extensive experiments have been run on five different control systems: vehicle turning, RLC circuit, DC motor position, aircraft pitch and quadrotor, each with three trials of sensor attacks: biased sensor value, sensor reading delay and sensor reading replay. 

Researchers have also taken this approach on specific applications, such as power grids overloading \cite{ma2020rmpc}. In power grids, the ultimate goal is to prevent overheating to break the system and cause destruction, and therefore the normal state is defined as small fluctuations in power generation and transmission. Quantitatively, the objective function of this recovery-based model predictive control methodology (RMPC) is to minimize generation ramping and load shedding with the presence of attacked components. A model predictive control problem, similar to Equation \eqref{eq:lp_recovery} but specifically for power grid load balancing, is formulated in their paper. The idea is the same: minimizing the objective under the constraints of system dynamics, control limits, and other desirable system properties.

\edit{\subsection{Machine Learning-based Control Synthesis}}
\label{subsec:black_box}
When the dynamics is unknown, solving for the recovery controller usually requires statistical inferences from observations. Specifically, machine learning techniques are introduced to solve for controls in this scenario. We identify two major approaches to devise a recovery controller in black-box dynamics: system identification and reinforcement learning.

The following paper presents an approach to first identify the system dynamics by statistics, and then apply an MPC controller. A subsequent work to \cite{ma2020rmpc} discussed in Section \ref{subsec:white_box}, the paper \cite{ma2021drmpc} considers a similar scenario to minimize thermal fluctuation in attacked power grids, except that the system dynamics is only partially known. The assumption here is that the dynamics is a linear time-invariant (LTI) system, but the coefficient matrices are unknown. Consequently, the first step is to perform system identification on the coefficients by training on input and output data when the CPS is attack-free, such as by regularized least squares method \cite{rifkin2003regularized}. Then, the same steps as in \cite{ma2020rmpc} on the surrogate dynamics, while the model parameters can be constantly updated during runtime by trusted data points. This procedure is denoted by the authors as recovery-based data-driven MPC (RDMPC).

The authors of \cite{ma2021drmpc} compare their RDMPC approach to a given dynamics method as in \cite{ma2020rmpc} by experiments, showing the capability of the learned model to correctly recover power overload even under a wrong assumption of system topology. 
These experiments are simulated on the IEEE test cases, which approximates the American Electric Power system \cite{ieee118bus}. In the upper subfigure, line 22 of IEEE 30-bus system is under attack and the RMPC with correct topology successfully recovers the system by maintaining the the current within its limit of 0.16 p.u. However, the given model in RMPC fails when the processor incorrectly believes line 27 is compromised. On the other hand, even with the wrong topology, the data-driven approach, RDMPC, still achieves the recovery objective. Similar results are shown in the bottom subfigure, where RDMPC resolves power overload in line 22 of IEEE 118-bus system under the false belief that line 15 and 20 are suffering denial-of-service. For more experiment details, please refer to the original paper.


\edit{
Another publication leverages sequence-to-sequence (Seq2Seq) models to predict the recovery control signal in a period of time \cite{liu2023seq}, under sensor attacks. Two Seq2Seq models are trained, each consisting of one encoder and one decoder. The first model serves as the state estimator to predict the actual physical state measurements from the corrupted sensor, while the second model serves as the recovery controller to output control signals from the estimated states. With this formulation, the recovery problem is reduced to a two-step multivariate forecasting problem, with the Seq2Seq models available off-the-shelf. For example, RNNs, LSTMs and GRUs \cite{medsker2001recurrent} can all be applied to the encoder-decoder implementation.
}

\begin{figure}
    \centering
    \includegraphics[width=0.8\textwidth]{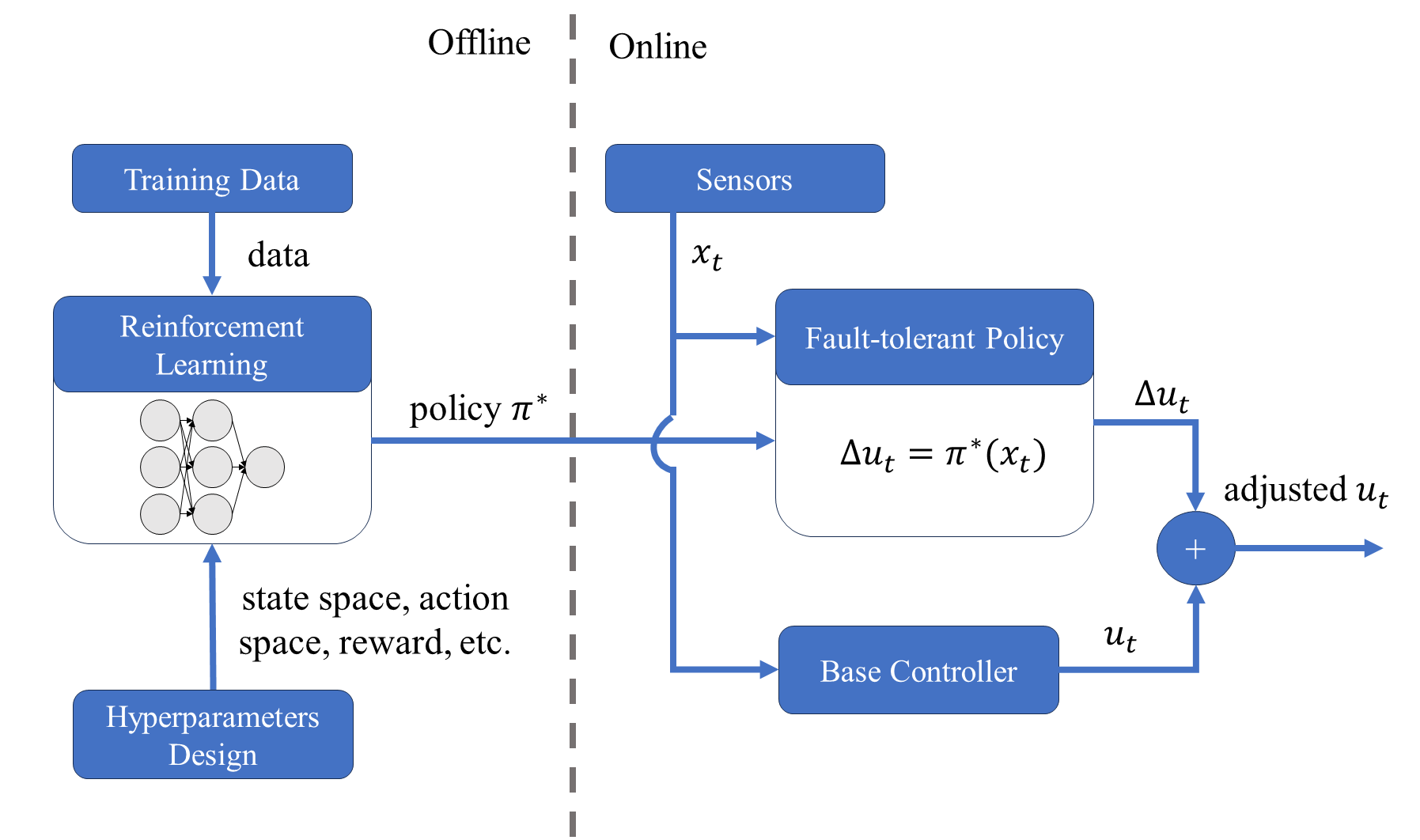}
    \caption{The reinforcement learning approach proposed by \cite{fei2021rl}, where an optimal control policy is learned offline and helps adjust the system's behavior at runtime.}
    \Description{Workflow of reinforcement learning-based recovery. At training phase, a control policy is learned, and at runtime, the policy is used to produce an additive control signal to a base controller.}
    \label{fig:reinforcement_learning}
\end{figure}

Different from using a surrogate dynamic plus a conventional controller, researchers in reinforcement learning has leveraged a purely learned approach \cite{fei2021rl}. Reinforcement learning is a machine learning procedure that learns a policy, i.e., what action to select at a given state to maximize not only the current reward, but also a potential future value \cite{watkins1992q, szepesvari1999unified, littman2001value, van2016deep}. It has already been proven useful in tasks such as chess, go, and competitive video games \cite{silver2018rlgames, chen2016alphago, shao2019rlVideoGames}, due to the fact that these tasks require strategies to maximize an overall score. In CPS recovery, we can see the process of reinforcement learning as an implicit system identification, such that the system is modeled as a Markov decision process, but instead of outputting a surrogate model, it directly produces a recovery control policy. In this paper, the policy takes in the current state and outputs an adjustment in the control signal for a base controller, e.g., PID controller. The adjustment aims to guide the system back to, or maintain the system inside a pre-defined normal state of high reward. The assumption is the same as any other reinforcement learning problem: a known state space, action space and reward function, with the rewards at each state denotes whether the state is normal or abnormal. The authors formalize a finite state space and an action space for a quadcopter example, and a reward function is designed as $r_t = ((f_{s_t} + f_{c_t})^2 + \epsilon_f)^{-1}$, where $f_{s_t}$ is the stability cost in form of a linear combination of state errors from references, $f_{c_t}$ is the control cost combining the control signal and its derivative, and $\epsilon_f$ is a slack variable. In other words, a normal or desirable behavior of the quadcopter is defined as flying steadily on its planned trajectory. Then, the training is done by actor-critic algorithm \cite{lillicrap2015actorCritic}. Finally, the learned fault-tolerant policy $\pi^*$ runs alongside the default PID controllers and adjusts their control signals based on state received. One significant advantage of this approach is the direct output of an optimal policy that is aware of a desirable behavior norm, but a drawback is that reinforcement learning requires large amount of training data and the careful design of state space and action space before runtime \cite{zhang2018rlOverfitting, wang2020rlControl}. The workflow of this approach is illustrated in Figure \ref{fig:reinforcement_learning}.


Different from \cite{fei2021rl}, the paper \cite{liu2023fulfilling} leverages reinforcement learning to directly compute a controller instead of assisting an existing controller. This dedicated recovery controller aims to guide a CPS back to a predefined safety target, specified by a signal temporal logic (STL) \cite{maler2004monitoring}, in shortest time possible. That is, the objective is to reach the goal as soon as possible (ASAP), which is a reasonable objective for real-time systems that requires fast responses. For example, a vehicle needs to park on the road shoulder ASAP when it suddenly has a flat tire. Under black-box dynamics, the solution framework, namely ASAP-Phi, leverages the following reward function
\begin{equation}
    \label{eq:asap_phi}
    r_{ASAP}(\bar{x}, \varphi, t) = 
    \begin{cases}
        QS(\bar{x}, \varphi, t) \text{ if } (\bar{x}, t) \not\models \varphi\\
        r >> \max_{\bar{x}', t'}{QS(\bar{x}', \varphi, t')} \text{ otherwise}
    \end{cases}.
\end{equation}
Here, $\bar{x}$ denotes a trajectory of states, i.e., $\bar{x} = x_0x_1\dots x_T$ for some finite $T \geq 0$. The target STL is denoted as $\varphi$. For example, $\varphi = x_t \in \mathcal{X}_{safe}$. A trajectory $\bar{x}$ can be evaluated at any time step $t \leq T$ on $\varphi$, and the outcome can be either true, meaning $\bar{x}$ at time $t$ satisfies $\varphi$, denoted as $(\bar{x}, t) \models \varphi$, or false, denoted as $(\bar{x}, t) \not\models \varphi$. Moreover, the function $QS$ denotes quantitative semantics \cite{hamilton2022training}, a finite real-valued measurement of the degree of satisfaction. The larger $QS(\bar{x}, \varphi, t)$ is, a larger degree that $\bar{x}$ at time $t$ satisfies $\varphi$. This can be seen as an extension from binary satisfaction outcomes to real numbers. As shown in Equation \eqref{eq:asap_phi}, this reward function $r_{ASAP}$ uses quantitative semantics scores when the trajectory is not yet fulfilling the target, and uses a large constant $r$ when it fulfills. The authors prove that this reward function is able to encourage fast satisfaction of a specification $\varphi$ under black-box dynamics, and it is insensitive of algorithms. Particularly, this reward function can be used along with any actor-critic-based \cite{peters2005natural} model-free reinforcement learning algorithms.

%% file: sections/5_exploratory.tex
\section{Exploratory Research}
\label{sec:exploratory}
There exist a few papers that do not directly purpose CPS recovery techniques. Instead, they discuss methods to assist recovery solutions. We therefore believe these are also worthwhile for review. Specifically, we find two papers that respectively consider two research problems: (1) order of recovery upon a mass failure on multiple CPS components, and (2) the adversary's intention upon an attack. By solving these two problems, we would have more information in constructing CPS recovery techniques.

\subsection{Recovery Order upon Mass Failure}
\label{subsec:recovery_order}

To decide the order of significance when recovering multiple failed components, and the authors of \cite{zhao2016interdependent} has provided a solution. Two models are therefore proposed based on different assumptions of the component-level recovery method: (1) reserved model (RM), where recovery is "prepaid", i.e., resources must be committed in advance of recovery and (2) opportunistic model (OM), where recovery is "pay-as-you-go", i.e., recovery can start as soon as some resources are available. At offline, each component $i$ is preassigned with a utility value $b^{(i)}$. \edit{The utility value of a component is a design choice, and the authors of this paper assumes same utility for all components in experiments.} Both models are then formulated as integer linear programming (ILP) problems that search for recovery orders that optimize some objective functions. Hence, the RM method can be reduced to
\begin{equation}
    \label{eq:recovery_order_ilp}
       \text{maximize } \sum_{i \in D}\sum_{t=1}^{T}b^{(i)}v^{(i)}_t\\
\end{equation}
where $D$ is the set of failed devices, $v^{(i)}_t$ is a binary variable indicating whether device $i$ has finished repairing at time $t = 1\dots T$. This RM ILP has three categories of constraints:
\begin{enumerate}
    \item Device dependency: a device can be repaired only if its supportive devices are working,
    \item Time for repair: a failed device returns to functioning only after a period of recovery time, and
    \item Sufficient resources for repair: a device can be repaired only if it has enough resources.
\end{enumerate}
Similarly, in OM, the same objective function is to be maximized but slightly different constraints shall be obeyed: 
\begin{enumerate}
    \item The "time for repair" constraint is loosened because recovery begins once part of the resources are allocated.
    \item The "sufficient resources for repair" constraint is also loosened due to the same reason.
    \item Moreover, additional constraints are added to this category to ensure recovery begins once resources are within a threshold.
\end{enumerate}
The ILP can be solved efficiently by algorithms such as dynamic programming. Please refer to the original paper for the mathematical formulation of these constraint functions.

The authors of \cite{zhao2016interdependent} present an experiment by simulating 50 failed devices in a general CPS, with each device assigned with a utility, maximum resources consumed at a time step (stage), and maximum number of stages for recovery. There is also a pre-defined amount of total resources available per stage. The experiment compares RM and OM algorithms, with the ILP solved by either an ILP solver or dynamic programming (DP). 


\subsection{Intention Interpretation}
\label{subsec:recovery_suggestions}

One interesting approach to suggest recovery actions is to infer what the attacker wants, assuming the system is under attack.
By knowing the intention of the adversary, the CPS is able to take countermeasures by intentionally avoiding the adversary's goal. For instance, if the attacker's purpose is found to be driving an autonomous vehicle to a specific coordinate, the system can see that location as a danger zone and modify its control to keep away from it.
The paper \cite{bezzo2018irl} approaches this problem with inverse reinforcement learning (IRL) \cite{ng2000irl}. Formally, IRL is to backward infer a reward function from a known control policy, and different from the healthy CPS's control policy $\pi$, the attacked CPS has an alternative policy $\pi' \neq \pi$ to guide the system behavior, which can be observed by traversing the state space. 



Therefore, the IRL problem is formalized as estimating the reward function that conveys the attacker's intention based on $\pi'$. At lower level, the reward function can be approximately solved by an active exploration method, which is updating the maximum likelihood estimation of the probability of a reward configuration given an observation, while traversing through the states, leveraging the Monte Carlo Markov Chain algorithm \cite{brooks2011mcmc}. Although not providing an explicit recovery technique, a well-estimated intention of the attacker facilitates designing counter actions, such as setting a penalty for going to the states highly rewarded by the attacker.

%% file: sections/6_discussion.tex
\section{Discussion}
\label{sec:discussion}

\subsection{Assumptions of Attacks}
\label{subsec:attack_types}

\begin{table}[ht!]
\setlength{\tabcolsep}{8pt}
\begin{center}
\caption{Attack types assumed by shallow recovery papers (Section \ref{sec:shallow}).
}
\vspace*{-5pt}
\label{tab:shallow_attacks}
{

\resizebox{\textwidth}{!}{
\begin{tabular}{r @{\hskip 7pt} cccccccccccccccc }

&
\rota{\cite{shin2019rnnDetector} Shin \emph{et al.}}&
\rota{\cite{mertoguno2019physics} Mertoguno \emph{et al.}} & \rota{\cite{loh2023enhancing} Loh and Thing} & \rota{\cite{lim2014arima} Nower \emph{et al.}} & \rota{\cite{wang2022security} Wang \emph{et al.}} & \rota{\cite{liu2022consensus} Liu \emph{et al.}} & \rota{\cite{wu2022integrated} Wu \emph{et al.}} & \rota{\cite{sheng2023collaborative} Sheng \emph{et al.}} & \rota{\cite{fawzi2014secure} Fawzi \emph{et al.}} &
\rota{\cite{yong2015resilient} Yong \emph{et al.}}& \rota{\cite{liu2016dynamic} Liu \emph{et al.}}& \rota{\cite{kong2018checkpointing} Kong \emph{et al.}}& \rota{\cite{choi2020software} Choi \emph{et al.}}& \rota{\cite{kong2021learning} Akowuah \emph{et al.}}&
\rota{\cite{alfaruque2019gan} Vatanparvar \emph{et al.}}&
\rota{\cite{kholidy2021autonomous} Kholidy}\\ 

\toprule 

\textbf{Attack surface} &&&&&&&&&&&&&&&&\\

Sensors &$\checkmark$ &&&&&&&& $\checkmark$ & $\checkmark$ & $\checkmark$ & $\checkmark$ &$\checkmark$&$\checkmark$&$\checkmark$&\\

Communication &&&& $\checkmark$ & $\checkmark$ & $\checkmark$ & $\checkmark$ & $\checkmark$ &&&&&&&&\\

General && $\checkmark$ & $\checkmark$ &&&&&&&&&&&&&$\checkmark$\\

\midrule

\textbf{Attack type} &&&&&&&&&&&&&&&&\\

DoS &&&& & $\checkmark$ & $\checkmark$ &&&&&&&&&&\\

Switch location &&&&&&&&&& $\checkmark$ & $\checkmark$ &&&&&\\

General & $\checkmark$ & $\checkmark$ & $\checkmark$ & $\checkmark$ && & $\checkmark$ & $\checkmark$ & $\checkmark$ &&& $\checkmark$ & $\checkmark$ &$\checkmark$ &$\checkmark$ &$\checkmark$\\
 
\bottomrule
\end{tabular}}}
\end{center}
\end{table}

\begin{table}[ht!]
\setlength{\tabcolsep}{8pt}
\begin{center}
\caption{Attack types assumed by deep recovery papers (Section \ref{sec:deep}).}
\vspace*{-5pt}
\label{tab:deep_attacks}

\begin{tabular}{r @{\hskip 7pt} ccccccccccccc }

&
\rota{\cite{abad2016restart} Abad \emph{et al.}}&
\rota{\cite{abdi2018restart} Abdi \emph{et al.}} & \rota{\cite{wang2022kinetostatic} Wang \emph{et al.}} & \rota{\cite{bayon2022cooperative} Bayon \emph{et al.}} & \rota{\cite{zhang2020lpRecovery} Zhang \emph{et al.}} & \rota{\cite{zhang2021lqrRecovery} Zhang \emph{et al.}} & \rota{\cite{zhang2023real} Zhang \emph{et al.}} & \rota{\cite{ma2020rmpc} Ma \emph{et al.}} & 
\rota{\cite{ma2021drmpc} Ma \emph{et al.}} &
\rota{\cite{liu2023seq} Liu \emph{et al.}} &
\rota{\cite{fei2021rl} Fei \emph{et al.}}& 
\rota{\cite{liu2023fulfilling} Liu \emph{et al.}}\\ 

\toprule 

\textbf{Attack surface} &&&&&&&&&&&&\\

Sensors &&&&&$\checkmark$&$\checkmark$&$\checkmark$&&&$\checkmark$&&&\\

General &$\checkmark$&$\checkmark$&$\checkmark$&$\checkmark$&&&&$\checkmark$&$\checkmark$&&$\checkmark$&$\checkmark$\\

\midrule

\textbf{Attack type} &&&&&&&&&&&&\\

General &$\checkmark$&$\checkmark$&$\checkmark$&$\checkmark$&$\checkmark$&$\checkmark$&$\checkmark$&$\checkmark$&$\checkmark$&$\checkmark$&$\checkmark$&$\checkmark$\\

\bottomrule

\end{tabular}
\end{center}
\end{table}

For shallow recovery papers listed in Table \ref{tab:shallow_attacks}, among the methods that uses redundant components, \cite{shin2019rnnDetector} considers a specific application on sensor redundancy and therefore assumes attacks on sensors. Feedback signal repair methods \cite{lim2014arima,wang2022security,liu2022consensus,wu2022integrated,sheng2023collaborative} are designed for information loss during communication, and assume attacks over communication channels. State estimation methods \cite{fawzi2014secure,yong2015resilient,liu2016dynamic,kong2018checkpointing,choi2020software,kong2021learning,alfaruque2019gan} assume corrupted sensors and therefore the measured state needs to be estimated. The remaining papers do not assume a particular attack surface. Moreover, except for \cite{wang2022security} and \cite{liu2022consensus} are specifically designed for DoS attacks, and \cite{yong2015resilient} and \cite{liu2016dynamic} are designed for switch location attacks, researchers in general do not care about the attack types.

For deep recovery papers listed in Table \ref{tab:deep_attacks},  \cite{zhang2020lpRecovery,zhang2021lqrRecovery,zhang2023real,liu2023seq} consider attacks on sensors, while the remaining papers do not care about the attack surface. Also, all these papers do not care about the attack type. In summary, the surveyed deep recovery papers only consider general faulty scenarios where the CPS needs to be guided back to normal, regardless of the type of attacks.

\subsection{Assumptions of the Dynamics}
\label{subsec:dynamics_knowledge}
The papers surveyed also assume for different types and different levels of knowledge of the dynamics. For example, we divide the deep recovery methods into recovery in white box and black box dynamics in Section \ref{sec:deep}. However, there exist more detailed categorization of the dynamics assumptions, and we tabulated them in Table \ref{tab:shallow_dynamics} and \ref{tab:deep_dynamics}.

\begin{table}[ht!]
\setlength{\tabcolsep}{8pt}
\begin{center}
\caption{Dynamics assumed by shallow recovery papers (Section \ref{sec:shallow}).
}
\vspace*{-5pt}
\label{tab:shallow_dynamics}
{

\resizebox{\textwidth}{!}{
\begin{tabular}{r @{\hskip 7pt} cccccccccccccccc }

&
\rota{\cite{shin2019rnnDetector} Shin \emph{et al.}}&
\rota{\cite{mertoguno2019physics} Mertoguno \emph{et al.}} & \rota{\cite{loh2023enhancing} Loh and Thing} & \rota{\cite{lim2014arima} Nower \emph{et al.}} & \rota{\cite{wang2022security} Wang \emph{et al.}} & \rota{\cite{liu2022consensus} Liu \emph{et al.}} & \rota{\cite{wu2022integrated} Wu \emph{et al.}} & \rota{\cite{sheng2023collaborative} Sheng \emph{et al.}} & \rota{\cite{fawzi2014secure} Fawzi \emph{et al.}} &
\rota{\cite{yong2015resilient} Yong \emph{et al.}}& \rota{\cite{liu2016dynamic} Liu \emph{et al.}}& \rota{\cite{kong2018checkpointing} Kong \emph{et al.}}& \rota{\cite{choi2020software} Choi \emph{et al.}}& \rota{\cite{kong2021learning} Akowuah \emph{et al.}}&
\rota{\cite{alfaruque2019gan} Vatanparvar \emph{et al.}}&
\rota{\cite{kholidy2021autonomous} Kholidy}\\ 

\toprule 

\textbf{Dynamics knowledge} &&&&&&&&&&&&&&&&\\

Known &&$\checkmark$&$\checkmark$&&$\checkmark$&$\checkmark$&&&$\checkmark$&$\checkmark$&$\checkmark$&$\checkmark$&&&&$\checkmark$\\

Unknown coefficients &&&&&&&&&&&&&$\checkmark$&&&\\

Unknown &$\checkmark$&&&$\checkmark$&&&$\checkmark$&$\checkmark$&&&&&&$\checkmark$&$\checkmark$&\\

\midrule

\textbf{Dynamics type} &&&&&&&&&&&&&&&&\\

Linear &&&&&&$\checkmark$&&&$\checkmark$&$\checkmark$&$\checkmark$&&$\checkmark$&&&\\

Nonlinear &$\checkmark$&$\checkmark$&$\checkmark$&$\checkmark$&$\checkmark$&&$\checkmark$&$\checkmark$&&&&$\checkmark$&&$\checkmark$&$\checkmark$&$\checkmark$\\
 
\bottomrule

\end{tabular}}}
\end{center}
\end{table}

\begin{table}[ht!]
\setlength{\tabcolsep}{8pt}
\begin{center}
\caption{Dynamics assumed by deep recovery papers (Section \ref{sec:deep}).
}
\vspace*{-5pt}
\label{tab:deep_dynamics}

\begin{tabular}{r @{\hskip 7pt} ccccccccccccc }

&
\rota{\cite{abad2016restart} Abad \emph{et al.}}&
\rota{\cite{abdi2018restart} Abdi \emph{et al.}} & \rota{\cite{wang2022kinetostatic} Wang \emph{et al.}} & \rota{\cite{bayon2022cooperative} Bayon \emph{et al.}} & \rota{\cite{zhang2020lpRecovery} Zhang \emph{et al.}} & \rota{\cite{zhang2021lqrRecovery} Zhang \emph{et al.}} & \rota{\cite{zhang2023real} Zhang \emph{et al.}} & \rota{\cite{ma2020rmpc} Ma \emph{et al.}} & 
\rota{\cite{ma2021drmpc} Ma \emph{et al.}} &
\rota{\cite{liu2023seq} Liu \emph{et al.}} &
\rota{\cite{fei2021rl} Fei \emph{et al.}}& 
\rota{\cite{liu2023fulfilling} Liu \emph{et al.}}\\ 

\toprule 

\textbf{Dynamics knowledge} &&&&&&&&&&&&\\

Known &$\checkmark$&$\checkmark$&$\checkmark$&$\checkmark$&$\checkmark$&$\checkmark$&$\checkmark$&$\checkmark$&&$\checkmark$&\\

Unknown coefficients &&&&&&&&&$\checkmark$&&&\\

Unknown &&&&&&&&&&&$\checkmark$&$\checkmark$\\

\midrule

\textbf{Dynamics type} &&&&&&&&&&&&\\

Linear &$\checkmark$&$\checkmark$&&&$\checkmark$&$\checkmark$&&$\checkmark$&$\checkmark$&&&\\

Nonlinear &&&$\checkmark$&$\checkmark$&&&$\checkmark$&&&$\checkmark$&$\checkmark$&$\checkmark$\\

\bottomrule

\end{tabular}
\end{center}
\end{table}














As shown in Table \ref{tab:shallow_dynamics}, shallow recovery papers with machine learning-based techniques do not require explicitly known dynamics. For example, \cite{shin2019rnnDetector} leverages RNNs to learn the system dynamics before sensor fusion, \cite{lim2014arima} uses ARIMA to interpolate missing measurements, and \cite{kong2021learning} utilizes LSTNet as a substitute of the dynamics model. On the other hand, techniques that does not leverage statistical inference (i.e. learning) of the dynamics require explicit models. The same situation applies to deep recovery papers as shown in Table \ref{tab:deep_dynamics}, where learning-based methods do not require the knowledge of dynamics while the others do. Moreover, \cite{choi2020software} and \cite{ma2021drmpc} assume known equation structure but unknown coefficients of the dynamics, and uses system identification techniques to infer the coefficients.  In both shallow and deep recovery, around half of the papers assume linearity of the dynamics, while the other half assume general, nonlinear dynamics.

\subsection{Additional Q\&A}
\label{subsec:q_and_a}
We also provide additional analysis in Q\&A form as follows.

\textbf{Q1: } Besides whether the recovery method is using a dedicated controller (i.e. shallow vs. deep recovery), what are some other dimensions to partition the recovery works?

\textbf{A1:} There are a few important dimensions that we identified. First, the specific CPS applications, such as power grids, sensor networks, robotics, etc., or unspecified. Second, the attack types, such as switch-location attack, false data injection, man-in-the-middle, etc., or unspecified. Third, the attack surfaces, such as sensors, controllers, actuators, communication networks, etc., or unspecified. Finally, we can divide the recovery scenarios into known system dynamics (white-boxes) vs. unknown system dynamics (black-boxes), which is already done in Section \ref{sec:deep}. Usually, a white-box dynamics implies the recovery method has guarantees in restoring the target physical state conditions, whereas black-box does not.

\textbf{Q2: } Based on the literature surveyed, what is the current state-of-the-art in CPS applications considered in the recovery research area?

\textbf{A2: } The majority of the surveyed works do not specify an application. Instead, they consider a general CPS consisting of sensors, controllers, actuators and optionally communication networks. Although some literature use an application as an example for their proposed method, the method can be trivially extended to other areas. For instance, \cite{lee2019rnn} uses velocity sensors on four-wheeled vehicles throughout the paper, but the component exclusion technique can be applied to any CPS with redundant components. However, a few of them are application-specific. For instance, \cite{liu2022consensus,ma2020rmpc,ma2021drmpc} consider power grids, with the recovery techniques particularly designed for power transmission objectives such as load balancing and frequency limits. Also, \cite{wang2022kinetostatic} focuses on quadruped robots, \cite{fei2021rl} on quadcopters and \cite{bayon2022cooperative} on an healthcare application of ankle exoskeletons.

\textbf{Q3: } Based on the literature surveyed, what is the current state-of-the-art in attack types considered in the recovery research area?

\textbf{A3: } Similar to application scenarios, most works do not specify an attack type. Instead, they focus on the type of consequences caused by the attacks. For example, the entire Section \ref{subsec:state_estimation} consists of solutions against loss of measured signals regardless of the attack type, but some of the papers (\cite{wang2022security} and \cite{liu2022consensus}) specifically focus on DoS attacks on communication networks. Similar situations happen in other sections, with two exceptions of \cite{yong2015resilient} and \cite{liu2016dynamic} consider switching location attacks. However, in the recovery wrapper provided by \cite{kholidy2021autonomous}, researchers have discussed various targeting recovery responses to particular attack types. In general, the recovery works focus on CPS malfunctions, and the solutions can be easily extended to various attack types when the malfunctions are the same.

\textbf{Q4: } Based on the literature surveyed, what is the current state-of-the-art in attack surfaces considered in the recovery research area?

\textbf{A4: } Two particular attack surfaces are heavily addressed by some of the papers: sensors and communication channels. Shallow recovery papers in Section \ref{subsec:state_estimation} (\cite{fawzi2014secure,kong2018checkpointing,yong2015resilient,liu2016dynamic,choi2020software,kong2021learning,alfaruque2019gan}) as well as two deep recovery papers in Section \ref{subsec:white_box} (\cite{zhang2020lpRecovery,zhang2021lqrRecovery,zhang2023real}) focus on sensor attacks. In this case, sensor readings are compromised due to attacks and therefore an estimation of the actual physical state shall be provided from a different information source. Meanwhile, the majority of papers in Section \ref{subsec:feedback_repair} (\cite{wang2022security,liu2022consensus,wu2022integrated,sheng2023collaborative}) focus on deprivation in communication channels, assuming different parts in the CPS are remote to each other. Therefore, when the communication signal is lost or corrupted, recovery strategies need to figure out the missing packets as well as optimize the CPS's actions for utilities. The remaining of the papers surveyed do not have a particular focus on attack surfaces. Instead, they counteract erroneous behaviors regardless of from what surfaces the attacks are injected.

\textbf{Q5: } Based on the literature surveyed, what is the current state-of-the-art in white-box vs. black-box dynamics considered in the recovery research area?

\textbf{A5: } First, some recovery strategies are not affected by whether the system knows the dynamics or not. For instance, the redundant components introduced in Section \ref{subsec:redundancy} only have to exclude the suspicious components in their processes. However, the knowledge of system dynamics may affect most of the methods. For example, the literature using physical state estimation will be heavily dependent on this knowledge, since a white-box dynamics allows straightforward roll-forwarding. In Section \ref{subsec:state_estimation}, among the 7 papers on state estimation, 4 of them assume white-boxes while 3 of them assume black-boxes. Moreover, the knowledge of dynamics also affects the synthesis of a dedicated recovery controller. In Section \ref{sec:deep}, 7 out of 10 deep recovery papers assume white-boxes and 3 black-boxes. Notice that some recovery methods not only assume white-box dynamics, but also have stricter requirements. For example, \cite{zhang2020lpRecovery} and \cite{zhang2021lqrRecovery} assume linear dynamics and \cite{abad2016restart}, \cite{abdi2018restart} and \cite{liu2023fulfilling} assume existence of stability.

\textbf{Q6: } What are relevant research backgrounds that can contribute to CPS recovery research?

\textbf{A6: } First, researchers with embedded systems backgrounds are able to apply their expertise in CPS architecture designs that aim for high resilience as well as recovery against attacks. Second, scholars with security background may extend their knowledge in security into this field. However, they shall mark the difference between CPS and software systems. For instance, as mentioned in \cite{abad2016restart}, rebooting a node or component has been a long-established method to recover a software system from malicious attacks, but in CPS the condition for rebooting shall be very different. Third, control system researchers are encouraged to promote recovery with dedicated controllers, i.e., deep recovery, where the recovery problem reduces to a control problem. More sophisticated control objectives, algorithms and frameworks are more than welcome. Fourth, machine learning researchers shall be capable of the situations where statistical inferences are required. For example, when the dynamics is a black-box, so that the CPS agent must first undergo a training phase. When traditional control guarantees are not applicable, machine learning metrics such as adversarial robustness can serve well to demonstrate the recovery performance. Last but not least, we encourage researchers in various application fields to participate in this sprouting domain, including healthcare systems, robotics, power grids, household sensor networks, etc. Interdisciplinary efforts are necessary when the research is brought to application level.

\textbf{Q7: } Any suggestions to researchers in these relevant fields, with regard to future research directions in CPS recovery?

\textbf{A7: } We encourage researchers to attempt novel sub-domains in this field. For example, as mentioned above, works have been done on sensor and communication channel attacks, but seldom on controller and actuators. Also, various types of attacks and applications may potentially endow more efficient and more specialized recovery strategies with higher performance. We sincerely hope all the specific sub-categories can become not only clearly identified but also populated with papers.

%% file: sections/6_1_challenges_and_future.tex
\edit{
\section{Future Directions and Challenges}
\label{sec:future}

We identify six major future directions for CPS recovery research as well as their corresponding challenges.

\textbf{1. Recovery under different specific scenarios.} Before conducting any research, the first step is always identifying the problem. For CPS recovery, this step is to answer the question: \textit{Against what kind of attacks, and with what assumptions of the system dynamics we are designing a recovery technique, and why we are researching this scenario?} As diverse CPS are evolving for numerous real-world applications, with different applications having different potential failures and objectives. For instance, a smart grid must avoid overheating and fluctuating power flow, and an autonomous vehicle must stay in safe physical states. Moreover, Section \ref{subsec:attacks} points out that not only CPS come in increasing diversity, but also attacks vary in multiple aspects such as purposes, surfaces and targets. Unfortunately, as shown in Section \ref{subsec:attack_types}, majority of the current literature has only focused on general attacks and failures on certain attack surfaces. These recovery methods are not designed for specific attack types or applications. Also, Section \ref{subsec:dynamics_knowledge} shows that the assumptions of system dynamics can be further detailed. So far, literature has only assumed for crude categories of linear vs. nonlinear systems, without considering more specific types of dynamics, such as time-variant vs. time-invariant, Lyapunov stability, and degree of nonlinearity.

Therefore, we consider one future direction is to identify more specific categories of recovery assumptions. A vital challenge of this direction is to distinguish what scenario assumptions are worth researchers to explore recovery techniques - this can be for multiple reasons, such as closeness to reality, life criticality, wide commercial use, or benefits to subsequent research. For instance, assuming a fully known noiseless linear time-invariant (LTI) dynamic may be handy for soundness proofs, but this assumption can be too ideal and too restricted to be applied to our physical world, so some alternative scenarios might be worth more efforts. Another example is that all CPS recovery papers assume the recovery module itself is perfectly safe and doing exactly what people designed, never considering its own vulnerability, which could lead to undesirable outcomes in applications.

\textbf{2. Improving recovery soundness.} Once a scenario is clearly identified, we ask the most significant question for a proposed recovery tool: Does it work? Specifically, as CPS gradually playing more life-critical roles in society, so that the most important metric to qualify a recovery method is \textit{how well does it save the system and its surroundings from the potential risks caused by attacks.} This is indeed the motivation we split the literature into shallow and deep recovery, with the latter unambiguously addressing the importance of normal physical states. Whether the recovery module intentionally drives the system back to normal provides one way to qualitatively evaluate soundness, and we can intuitively assert that the deep recovery methods provide better soundness than the shallow ones. It is a pity that only a few papers so far fall into the category of deep recovery, while most remain shallow.

Nevertheless, one remaining challenge is to formalize soundness, preferably quantitatively, in a standardized way that researchers are able to agree on. For instance, an upper-bounded distance to a pre-defined safe set after recovery may convince people that this method is sound, with the bound itself a metric. Some methods, especially machine learning methods, might not be able to give deterministic guarantees due to randomness, so that probabilistic measures can serve as substitutes. With a well-defined quantitative soundness, a succeeding challenge is to design recovery methods that improve this metric at the trade-off with time and resource overheads. For example, a learned state estimator's soundness can be improved if the model approximates the underlying distribution better, but this improvement requires training on a larger or less biased data set. Consequently, to maintain training data overhead small in development phase, a novel data collection method for unbiased data points can contribute.

\textbf{3. Reducing latency for real-time systems.} CPS are usually real-time systems, and some have very strict requirements in fast reactions, i.e., low latency. For instance, drones and autonomous vehicles must interact with the environment at high frequency and react fast to sudden changes. Otherwise, property damages or lethal consequences may occur. Therefore, when designing recovery methods, two questions must be answered: \textit{(1) Will running the recovery module pause or slow down the system? (2) How long does it take to run the recovery, and if it takes a non-trivial period of time, is the computation result still useful as the situation may be different?} Again, the challenge will be the tradeoff between speed and other dimensions like soundness and memory consumption. Some recovery methods are fast, such as restarting discussed in Section \ref{subsec:white_box}, but there is low to no guarantee that these methods can fix the problem.

\textbf{4. Reducing resource consumption at both development phase and runtime.} We must consider economy in CPS, which can be burdened by tedious production procedures and high requirements for hardware at runtime. Therefore another question rises: \textit{How much does the recovery module cost during development and runtime?} Unlike traditional software systems, CPS are generally built as embedded systems, i.e., systems for a specific function, such as robotic arms and unmanned vehicles. Hence, CPS generally do not require complete operating systems nor complex architecture. Instead, light-weighted hardware and software with low developmental and runtime costs promote mass production. Particularly, a recovery module is a software that built on top of an existing CPS, and must not induce large resource consumption before and during runtime compared to other built-in processes.

As mentioned in the previous subsection, one large consumption at design phase shall be data collection and training for machine learning models. This is a reasonable concern because we would better not assume all information needed by recovery are given, and therefore some knowledge shall be inferred by statistics. Hence, balancing among learning model complexity, accuracy and data cost when producing a CPS recovery module becomes one research challenge. Another cost is the hardware resource overhead during runtime such as memory, CPU and network bandwidth, which also needs to be taken into consideration when devising recovery.

\textbf{5. Attack Diagnosis for Recovery.}
As noted, the binary result of attack detection, i.e. under attack or not, is insufficient.
The recovery also needs attack diagnosis to identify not only `who': which components are corrupted, but also `'when: when the attack starts.
Note that diagnosing the starting time is equally important since this will tell until when the historical data are trustworthy, which thus the recovery can rely on.
That is, after attack detection, the diagnosis is activated to find out the trustworthy data that the recovery can use.
Using compromised historical data may result in a failed recovery, e.g. driving a system to an unsafe state.
In spite of the importance, attack diagnosis has been inadequately addressed so far.
Consider recovery from sensor attacks as an example. 
To launch the appropriate recovery needs to know both the `who' and `when'.
Although some existing sensor attack detection methods may be extendable to find out which sensors are compromised~\cite{zhang2022adaptive,wang2021hada,akowuah2021realtime,quinonez2020usenix}, none can do when they start to be compromised.
For each sensor, we may set a threshold for the residual between the predicted and observed values.
Then, a sensor is considered as under attack if the residual is greater than its threshold. 
However, this cannot be used to determine when the sensor starts to be under attack, and using the detection time, i.e. the moment when a detector raises the alarm, is not accurate. 
This is because the attack detector may take some time to ascertain the occurrence of an attack.
Publication~\cite{wang2023attention} is considered as the first work to address this temporal sensor attack diagnosis problem in CPS.
The authors in~\cite{wang2023attention} develop an attention mechanism based attack diagnosis method, which utilizes the fluctuations of attention scores as an indicator of the start of an attack.
Here we call for more efforts on effective and efficient attack diagnosis methods that can real-time and accurately provide both the `who' and `when' to enhance the success of attack recovery.

\textbf{6. More exploratory research topics.}
Besides concrete recovery methods themselves, more exploratory research topics like the papers \cite{zhao2016interdependent,bezzo2018irl} in Section \ref{sec:exploratory} needs to be investigated. For instance, one example topic is to theoretically or empirically analyze the trade-offs among the dimensions stated above. Other topics that can possibly impact CPS recovery soundness or efficiency need also be analyzed. By answering these questions, researchers do not directly build CPS recovery frameworks; instead, they provide valuable information on which future recovery methods can be constructed.
}

%% file: sections/7_conclusion.tex
\section{Conclusion}
\label{sec:conclusion}

We defined the concept of CPS recovery as the restoration of physical states under adversarial attacks. Then, we surveyed the current state-of-the-art papers that satisfy our definition, with a total of \edit{30} papers thoroughly reviewed from year 2014 to 2023. Among these papers, we identify a major partition of not utilizing vs. utilizing a dedicated controller to perform a recovery task, which we denoted as shallow vs. deep recovery. Finally, in the discussion section, we discussed additional partitions of the literature, including applications, attack types, attack surfaces and knowledge of the system dynamics. Then, we analyzed the current state-of-the-art based on these partitions as well as provide suggestions for future research.

%% file: sections/acknowledgement.tex
\section*{Acknowledgement}

This work was supported in part by NSF CNS-2143274 and NSF CNS-2333980.
The views and conclusions contained herein are those of the authors and should not be interpreted as necessarily representing the official policies or endorsements, either expressed or implied, of the National Science Foundation (NSF).